%% file: main.tex
\documentclass[conference]{IEEEtran}

\usepackage{cite}

\ifCLASSINFOpdf
  \usepackage[pdftex]{graphicx}
\else
\fi

\usepackage{xurl}
\usepackage{amsmath}
\usepackage{xcolor}
\usepackage{booktabs}
\usepackage[hidelinks]{hyperref}
\usepackage{cleveref}
\usepackage[numbers, sort&compress]{natbib}
\usepackage{graphicx}
\usepackage[caption=false,font=footnotesize]{subfig}
\usepackage{colortbl}
\usepackage{multirow}
\usepackage[table]{xcolor}
\usepackage{array} 
\usepackage{makecell}   
\usepackage{booktabs}  
\usepackage{tabularx}  
\usepackage{multirow}  
\usepackage{ragged2e}  
\usepackage{multicol} 
\usepackage{enumitem} 
\usepackage{algorithm}
\usepackage{algpseudocode}
\usepackage{tcolorbox}
\usepackage{fontawesome5}
\usepackage{twemojis}

\usepackage{listings}
\usepackage{xcolor}

\lstdefinelanguage{HTML}{
  sensitive=true,
  keywords={html, head, body, div, span, img, a, href, src, style, class},
  morestring=[b]",
  morecomment=[s]{<!--}{-->},
  stringstyle=\color{teal},
  keywordstyle=\color{blue},
  commentstyle=\color{gray},
}

\begin{document}
\input{cmds}

%
\title{A Large-Scale Empirical Study of Modern Phishing Email Content}

\author{
\IEEEauthorblockN{Jaehwan Park\IEEEauthorrefmark{1},
Woonghee Lee\IEEEauthorrefmark{2},
Fujiao Ji\IEEEauthorrefmark{1},
Doowon Kim\IEEEauthorrefmark{1}}
\IEEEauthorblockA{\IEEEauthorrefmark{1}University of Tennessee, Knoxville \\
\{jpark127, fji1, doowon\}@utk.edu}
\IEEEauthorblockA{\IEEEauthorrefmark{2}Korea University \\
whlee@isslab.korea.ac.kr}
}


\maketitle

\begin{abstract}
Phishing remains one of the most pervasive threats to Internet users. Email remains its predominant delivery channel, although the attack vectors have expanded to SMS. Email content is the attack surface of phishing because it is what the victim reads and what automated defenses inspect. Yet the composition of modern phishing content is poorly measured. Although prior work has characterized dimensions such as theme, call-to-action (CTA), and impersonation, these dimensions have not been measured at scale, nor have their associations or temporal changes been systematically understood. These gaps stem from limitations in both data and method: prior work has relied on small, historical, or source-specific corpora; used bag-of-words topic models that provide limited semantic context; and focused primarily on textual email content.

We present a content-focused measurement study of  2.9M distinct real-world phishing emails collected over 13 months (June 2025 -- June 2026), in collaboration with the Anti-Phishing Working Group (APWG). We treat each email as a composite artifact comprising message text and the attachments it carries: 272K images, 143K PDFs, and 57K calendar invitations. Using an LLM pipeline validated against human-annotated samples, we analyze these components along three dimensions (theme, CTA, and impersonation). We further examine the associations among these dimensions, and we measure longer-term change by comparing our corpus against a historical dataset.

We find that attackers diversify what they use to deceive but converge on how victims should respond: no theme exceeds 21.3\% of emails, while a single CTA, URL navigation, accounts for 73.0\%. CTA and impersonation choices are systematically conditioned on theme. Attachments fall into three roles: \textit{supplement}, \textit{substitute}, and \textit{reinforce}. Images supplement the message text, PDFs carry the pretext in place of it, and calendar invitations replicate interaction endpoints into a persistent medium. Bridging our corpus with a historical dataset, we show that the dominant CTA for invoice-themed phishing shifted from URL navigation to offline communication: offline communication accounted for 6.7\% in 2015 but 46.9\% in 2025.
\looseness=-1

\end{abstract}

\input{tex/introduction}
\input{tex/background}
\input{tex/prblemstatement}
\input{tex/datacollection}

\input{tex/overviewOfMethod}
\input{tex/research_question_1_new_2}

\input{tex/research_question_3_new_3}
\input{tex/research_question_4_new_2}
\input{tex/new_related_work_2}

\input{tex/discussion_2}


\input{tex/conclusion_2}



\input{tex/openscience_2}

\bibliographystyle{IEEEtran}
\bibliography{ref.bib}

\input{tex/appendix_ndss_2}



\end{document}

%% file: cmds.tex
\newcommand{\DK}[1]{\textcolor{purple}{\bf [DK: #1]}} 
\newcommand{\FIXME}[1]{\textcolor{red}{\bf [FIXME: #1]}} 
\newcommand{\JP}[1]{\textcolor{orange}{\bf [JP: #1]}} 
\newcommand{\WL}[1]{\textcolor{magenta}{\bf [WL: #1]}} 
\newcommand{\FJ}[1]{\textcolor{blue}{\bf [FJ: #1]}} 

\newcommand{\PP}[1]{
\noindent{\bf {#1.}}}

\newcommand{\eg}{\textit{e}.\textit{g}.}
\newcommand{\ie}{\textit{i}.\textit{e}.}
\newcommand{\best}[1]{\cellcolor{gray!20}\textbf{#1}}

\renewcommand{\algorithmicrequire}{\textbf{Input:}}
\renewcommand{\algorithmicensure}{\textbf{Output:}}

\definecolor{lightgray}{gray}{0.9}

\newtcolorbox{boxI}{
    colback = lightgray!10, 
    colframe = black, 
    boxrule = 0.5pt, 
    toprule = 0.5pt, 
    arc = 2pt,
    left = 1pt,
    right = 1pt,
    bottom = 0pt,
    top = 0pt
}
\newcounter{observcntr}
\newcommand*{\observ}[1]{%
    \stepcounter{observcntr}%
    \begin{center}
        \begin{boxI}
        \textbf{Takeaway~\arabic{observcntr}: }{#1} %
        \end{boxI}
    \end{center}
}

\def\Appendixautorefname{\Appendix}
\renewcommand{\figureautorefname}{Figure}
\renewcommand{\tableautorefname}{Table}
\providecommand*{\listingautorefname}{Listing}
\renewcommand{\sectionautorefname}{Section}
\renewcommand{\subsectionautorefname}{Section}
\renewcommand{\subsubsectionautorefname}{Section}
\renewcommand{\subsubsectionautorefname}{Section}
\renewcommand{\Appendixautorefname}{Appendix}


%% file: tex/introduction.tex
\section{Introduction}
\label{sec:introduction}












Phishing remains one of the most pervasive threats facing Internet users \cite{fbi2025ic3report:online}.
Although the attack has expanded to other channels, including SMS, email remains the predominant delivery channel \cite{sun2024victims,lain2025url}.
In a typical phishing campaign, an attacker composes a deceptive message (\eg, an urgent password change notice), and pairs it with a payload (\eg, a link to a phishing site) that steers the recipient toward an attacker-controlled website for credential harvesting.

Email content is the attack surface of phishing. 
It is what the potential victim reads, what an automated defense inspects~\cite{verma2012detecting,fette2007learning}, and what awareness training teaches people to recognize~\cite{kumaraguru2007protecting}.
Therefore, understanding how modern phishing content is constructed is a prerequisite for improving all three.
Prior work, \textit{however}, offered only a limited landscape of this content, and its limitations fell along three axes: the \textit{dataset} it relied on, the \textit{method} it applied, and the \textit{unit of analysis} it adopted.
\textit{First (dataset)}, previous studies \cite{saka2024phishing,dalmiere2025measuring,longtchi2024characterizing} relied on small or historical datasets, which provide limited visibility into the contemporary phishing ecosystem.
This limitation is structural rather than incidental. 
Because phishing arrives in personal and corporate mailboxes, privacy concerns severely constrain both the collection of phishing corpora and their public release.
Moreover, prior work has relied on provider- or organization-specific datasets collected from a small number of organizations~\cite{simoiu2020targeted,van2019cognitive}, which limits generalization to the broader phishing landscape.


\textit{Second (method)}, in prior work~\cite{van2019cognitive,hoheisel2023development}, the characterizations rested on Latent Dirichlet Allocation (LDA)~\cite{blei2003latent} or Labeled LDA~\cite{ramage2009labeled}, a bag-of-words model that captures only lexical co-occurrence, but not the semantics that make a lure persuasive. 
For example, it cannot separate a message that impersonates a bank to threaten account suspension from one that impersonates the same bank to promise a refund, even though the two exert very different pressure on the recipient.

\textit{Third (unit of analysis)}, prior work was narrow in two respects.
It concentrated on the textual email content of the email message~\cite{hoheisel2023development,van2019cognitive}, overlooking the persuasive content that attackers place in attached images, PDFs, and calendar invitations.
These components travel in the same message and can carry the attack content itself~\cite{pdftalos1:online,barracuda2026calendar,sublimeimage1:online,dalmiere2025measuring}, yet they have received little systematic attention. 
Prior work has characterized phishing content along dimensions such as theme, requested action, and impersonation using small datasets~\cite{saka2024phishing,saka2024phishcoder}, but has not systematically quantified the associations among these dimensions at scale or their temporal change.


We address these gaps with a measurement study of 2.9M distinct real-world phishing emails collected over 13 months, from June 2025 to June 2026, in a collaboration with the Anti-Phishing Working Group (APWG)~\cite{apwgemailreport1:online}. 
We analyze each phishing email as a composite artifact, examining its message text together with the attachments it carries, including images, PDFs, and calendar invitations. 
We also characterize every component along three dimensions, namely theme, call to action (CTA), and impersonated entity, and we quantify how these dimensions co-occur, both within message text and across non-text components. 
To enable the analyses at scale, we leverage an LLM (GPT-5.1) as an annotator and validate its outputs against human-annotated samples.
Finally, we compare our contemporary dataset against a historical phishing corpus to examine how tactics within persistent themes have shifted over time.

Our work is guided by three research questions:
\begin{itemize}[leftmargin=*, itemsep=2pt, topsep=0pt]

  \item \textbf{RQ1 (Composition):} How is contemporary phishing content constructed along theme, CTA, and impersonated entity, and how do these three dimensions co-occur in practice?

  \item \textbf{RQ2 (Attachment):} What content do attached images, PDFs, and calendar invitations carry, and what role do they play in the campaign?

  \item \textbf{RQ3 (Temporal Change):} Within themes that persist across time, which CTA and impersonation patterns have changed, and which have remained stable?
\end{itemize}

\PP{Key Findings}
Addressing RQ1, we first characterize the three dimensions individually. 
We observe that themes are broadly distributed, with no single theme exceeding 21.3\%, whereas CTAs are highly concentrated on a single CTA (URL navigation, 73.0\%), and impersonation is prevalent but dispersed across a long tail of brands. 
This indicates that attackers diversify \emph{what} they use to deceive (\ie, various themes), but converge on \emph{how} victims should respond (\ie, visiting a phishing site). 

Examining associations across dimensions further shows that these dimensions are not independent.
Both the choice of CTA and the use of impersonation are conditioned on the attack theme (\autoref{sec:researchquestion1}).
For example, invoice-themed emails more often direct victims to phone calls than URLs, and extortion-themed emails commonly demand cryptocurrency payments.
Moreover, the use of impersonation follows the same logic.
It is common in themes (\eg, government and invoice) that borrow organizational legitimacy (or a recognizable sender), but is much rarer in themes (\eg, extortion and romance) that can operate without the legitimacy (see~\autoref{fig:rq1_theme_impersonation_none}).

Second, addressing RQ2, we find that the three attachment types play three distinct roles related to the body text, which we term 
\emph{supplement}, \emph{substitute}, and \emph{reinforce}.
Images \emph{supplement} the message text, supplying content the message cannot express in prose: document scans for invoices, human photographs for romance lures, and logos for impersonated brands.
PDFs instead \emph{substitute} for the message body text, carrying the pretext itself rather than supplementing it. Consistent with this role, emails carrying PDFs have substantially shorter message bodies than emails without PDFs. PDF content is also highly specialized: 64.6\% of all PDFs occur in a single configuration of the three dimensions---an Invoice theme, an Offline Communication CTA, and private-company impersonation.
Calendar invitations (ICS) \emph{reinforce} it: 97.8\% of phone-bearing ICS attachments repeat a phone number already present in the body text, replicating the contact endpoint into a persistent medium that resurfaces through calendar reminders.

Finally, addressing RQ3, we find that persistent themes follow divergent trajectories (\autoref{sec:researchquestion3_historical}).
The invoice theme shows the sharpest shift: Offline Communication accounted for 6.7\% of Invoice CTAs in 2015 and became the dominant CTA by 2025, reaching 46.9\% of Invoice CTAs.
By contrast, the Account theme retains a stable CTA structure, while its vocabulary shifts from mailbox administration (\eg, mailbox) to cloud-stored content (\eg, cloud, files, and storage).

Together, these results can give defenders an empirical account of the content adversaries actually deliver, and translate it into lessons for hardening the email ecosystem against the components that current defenses may under-examine (\autoref{sec:discussion}).



In summary, the contributions of our paper are as follows:
\begin{itemize}[leftmargin=*, itemsep=2pt, parsep=0pt, topsep=0pt, partopsep=0pt]
    \item We conduct the first large-scale, content-level measurement of modern phishing, characterizing 2.9M distinct real-world emails along three dimensions (theme, CTA, and impersonated entity), and quantifying their associations. This reveals that CTA and impersonation strategies are systematically theme-dependent.

    \item We present a large-scale paired analysis of message text and attachments, covering 272K images, 143K PDFs, and 57K calendar invitations. We show that attachments occupy an extend-replace-duplicate spectrum: images \textit{supplement} the text, PDFs \textit{substitute} for it, and calendar invitations \textit{reinforce} its endpoints into a more persistent medium, implying that text-only detection has little to inspect in a substantial fraction of emails.

    \item We place contemporary patterns in historical context by bridging our corpus with the Nazario corpus (2015–2025). We show that characteristics within the same persistent theme follow different temporal trajectories: Invoice's dominant CTA shifts to offline communication in 2025, while Account remains stable.

\end{itemize}

%% file: tex/background.tex
\section{Background}
\label{sec:background}
\begin{figure}[t]
    \centering
    \includegraphics[width=0.8\linewidth]{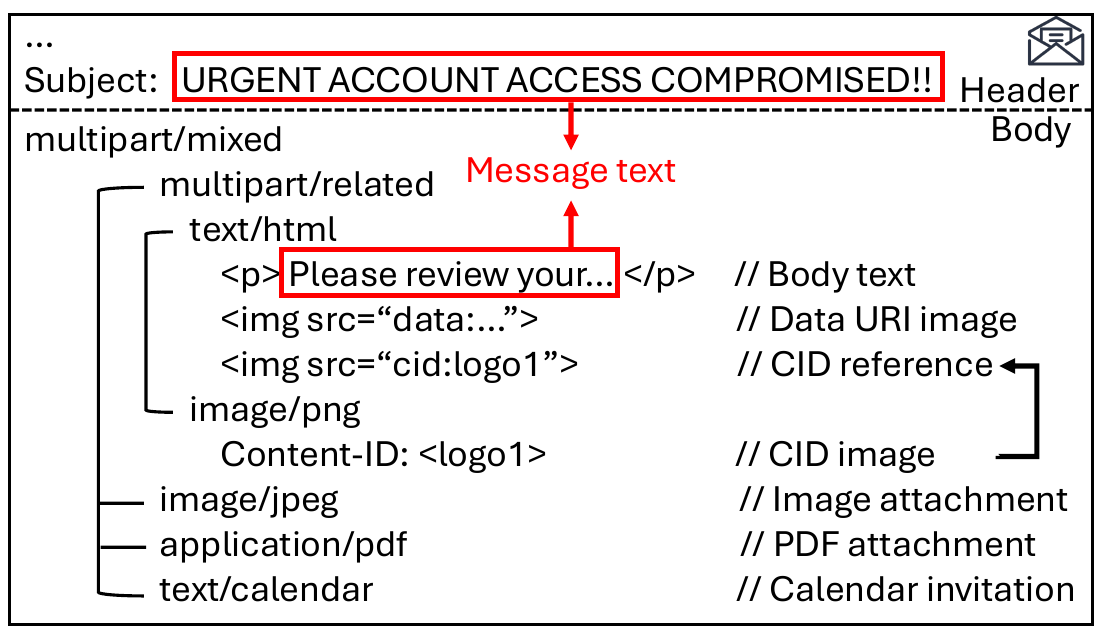}
    \caption{\textbf{Example of MIME Structure.}}
    \label{fig:background_mime_example}
\end{figure}

\PP{Phishing Attacks}
Phishing attackers use social engineering to obtain credentials, money, or sensitive information (\eg, Social Security numbers)~\cite{statistics3}.
Phishing can cause direct financial losses and serve as an access vector for broader attacks against organizations~\cite{yuldoshkhujaev2025decade,fbi2026kimsukhi:online}.
Email remains a prevalent delivery channel for phishing attacks~\cite{moore2007examining,verma2012detecting}.
In phishing emails, attackers present content, request actions from recipients, and may impersonate trusted organizations (\eg, Facebook).


\PP{Phishing Email Structure}
Internet email consists of a header and a message body~\cite{rfc5322:online}, while MIME allows the body to contain multiple content parts and attachments~\cite{rfc2045:online,rfc2046:online}. \autoref{fig:background_mime_example} illustrates an example MIME structure.
Images can be embedded directly in an HTML body through data URIs~\cite{rfc2397:online}, referenced as MIME parts through content identifiers (CIDs)~\cite{rfc2392:online}, or attached to the email body~\cite{rfc2183:online}.
PDFs and calendar invitations can appear as MIME parts or attachments.
In this work, we refer to the subject line and visible body text collectively as message text.

\PP{Content Across Email Components}
This structure allows phishing content to appear in different parts of the same email. The message text may contain the phishing lure and requested action directly, or it may provide only limited context while an image, PDF, or calendar invitation carries additional attack content. Therefore, analyzing only the message text does not capture how phishing content is distributed across an email. We examine these components together and relate their content to the surrounding message text.

%% file: tex/datacollection.tex
\section{Analysis Plan for Phishing Email Ecosystem}
\label{sec:method}

\begin{figure}[t]
    \centering
    \includegraphics[width=\linewidth]{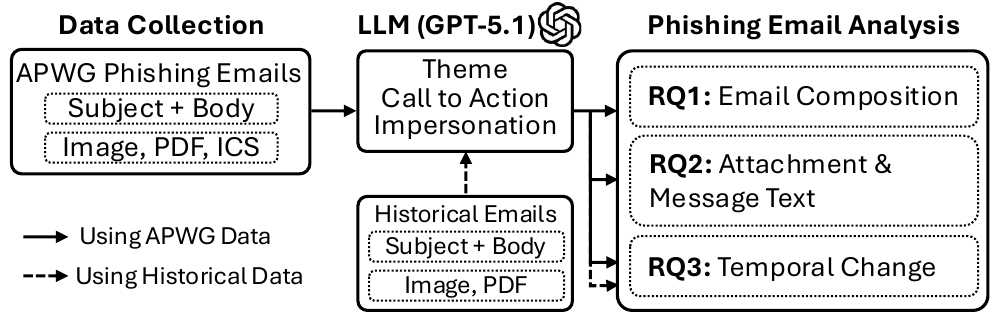} 
    \caption{\textbf{Overview of Phishing Email Analysis.} We examine (1) themes, call-to-action, and impersonation in message text; (2) phishing content in images, PDFs, and calendar invitations; and (3) temporal changes.}
    \label{fig:overview}
\end{figure}

\PP{Overview}
As illustrated in~\autoref{fig:overview}, our methodology proceeds in three stages: (1) we collect 3.8M real-world phishing emails over 13 months, from June 2025 to June 2026, through a partnership with the Anti-Phishing Working Group (APWG)~\cite{apwgemailreport1:online}.
(2) We refine the collected emails (\eg, removing duplicates) and extract their subjects, message bodies, images, and attachments, resulting in 2.9M emails for analysis.
(3) We employ GPT-5.1~\cite{gpt51:online} to analyze the extracted content at scale, as manually examining 2.9M emails is almost infeasible.

Our analysis addresses three research questions.
(1) \textit{How is modern phishing content constructed along theme, CTA, and impersonation, and do these dimensions co-occur in practice?}
We characterize message text through phishing themes, call-to-action (CTAs), and impersonation, and examine the associations between these dimensions.
(2) \textit{What phishing content appears in images, PDFs, and calendar invitations, and how does it relate to the surrounding message text?}
We examine themes, CTAs, and impersonation in these components and compare them with the surrounding message text.
(3) \textit{How have persistent phishing themes changed over time?}
We examine whether the patterns observed in our dataset are recent or have persisted over time within the same themes.

\subsection{Real-world Phishing Email Collection}
\label{sec:method:data:collection}
We first collect a large-scale dataset of real-world phishing emails through a partnership with the Anti-Phishing Working Group (APWG), a global coalition of industry stakeholders (\eg, Microsoft, RSA Security, Verisign, PayPal, etc.). The APWG facilitates a centralized reporting ecosystem where members submit suspected phishing emails, which are subsequently reviewed and validated by the consortium. This validation process yields a high-fidelity dataset distinct from the noise often found in crowdsourced collections~\cite{apwgemailreport1:online}.
\looseness=-1

\begin{table}[t]
\caption{\textbf{Overview of Our Collected Dataset.}
}
\label{tab:dataset}
\centering
\footnotesize
\begin{tabular}{l r}
\toprule
    \multicolumn{1}{c}{\textbf{Type}} & \multicolumn{1}{c}{\textbf{\# of Emails}} \\
     \midrule
  Total Reported Phishing Emails & 3,792,557 (100.0\%)\\
  \midrule
  
\textbf{Refined Total Emails} & 2,898,398~\phantom{0}(76.4\%)\\
  
   $\rightarrow$ Email without Attachments  & 2,526,736~\phantom{0}(66.6\%)\\
  $\rightarrow$  Email with Attachments & 371,662~\phantom{0}\phantom{0}(9.8\%)\\
\phantom{0} $\hookrightarrow$  Emails with Image & 188,821~\phantom{0}\phantom{0}(5.0\%) \\
\phantom{0} $\hookrightarrow$ Emails with PDF & 138,992~\phantom{0}\phantom{0}(3.7\%) \\
\phantom{0} $\hookrightarrow$ Emails with ICS & 56,831~\phantom{0}\phantom{0}(1.5\%) \\

\midrule
\midrule
Collection Period & 06. '25 -- 06. '26 (13 Months) \\
\bottomrule
\multicolumn{2}{l}{\scriptsize\textit{Note: Percentages in this table use the 3,792,557 reported emails as the denominator;}} \\
\multicolumn{2}{l}{\scriptsize\textit{ subsequent analyses use the 2,898,398 refined emails.}} \\
\end{tabular}
\end{table}




\PP{Collection and Refinement of Dataset}
We collect a total of 3.8 million phishing emails over 13 months (Jun. `25 -- Jun. '26). 
To ensure data quality and consistency, we refine the raw dataset by removing duplicates (\ie, emails with identical contents) and structurally incomplete emails (\eg, those with no message body).
As shown in~\autoref{tab:dataset}, these refinement steps yield 2.9M valid phishing emails (76.4\% of the total 3.8M).


\PP{Parsing and Extraction}
From each email, we extract two types of content, which we refer to throughout
the paper as \emph{message text} and \emph{attachments}. Message text is the
subject line together with the visible body text. Attachments are content
preserved with the email itself: (1) images embedded as data URIs in
HTML~\cite{rfc2397:online}, CID-referenced images carried as MIME
parts~\cite{rfc2392:online}, standalone image
attachments~\cite{rfc2183:online}, (2) PDFs\footnote{We found a single PDF embedded as a data URI in the corpus and we did not separate it in our analysis.}, and (3) calendar invitation (ICS) files.

\begin{itemize}[leftmargin=*, itemsep=2pt, parsep=0pt, topsep=0pt, partopsep=0pt]
    \item \textbf{Message Text}: We parse HTML bodies with BeautifulSoup4~\cite{beautifu17:online}. For \texttt{multipart/alternative} messages we use the HTML representation when one is available, since clients render it in preference to the plain-text alternative~\cite{rfc2046:online}. Because our analysis targets the content a recipient (\ie, potential victim) actually sees, we discard text hidden from view: phishing emails routinely embed invisible text to evade detection~\cite{salttalos1:online}. Specifically, following WebAIM accessibility guidance~\cite{webaim1invisibe:online}, we remove elements carrying the \texttt{hidden} attribute or a \texttt{display:}~\texttt{none} style. We analyze the invisible text in Appendix~\autoref{appendixsec:hidden_texts}.

    \item \textbf{Attachment}: We extract PDF, image, and calendar-invitation attachments from the MIME structure using Python's \texttt{mail-parser} library~\cite{mailparser:online}
    , decoding each part according to its Content-Transfer-Encoding. 
    This yields 371,662 emails containing at least one attachment (see~\autoref{tab:dataset}), including 188,821 with images, 138,992 with PDFs, and 56,831 with ICS attachments.
\end{itemize}

\PP{Data Sanitization for Privacy}
To mitigate privacy and ethical concerns when analyzing the email dataset through commercial LLMs (in this paper, GPT), 
we remove information introduced through the reporting process that could identify phishing reporters or recipients, including recipient email addresses and other reporting metadata not required for our analysis.
We submit only the email content required for classification.
Furthermore, OpenAI states that data submitted through its API is not used to train its models by default~\cite{openapi1:online}.
API data may be retained in abuse-monitoring logs for up to 30 days, subject to limited exceptions~\cite{openapi1:online}.

\begin{table}[t]
\caption{\textbf{Performance Comparison of LLMs on Message Text.}
We report support-weighted precision, recall, and F1.}
\centering
\small
\setlength{\tabcolsep}{3pt}
\resizebox{\columnwidth}{!}{%
\begin{tabular}{llccccc}
\toprule
\textbf{Task} & \textbf{Metric}
& \textbf{GPT-5.1}
& \textbf{GPT-5~mini}
& \textbf{Gemini-3.0$^1$}
& \textbf{Qwen3-8B}
& \textbf{L3.1-8B-I$^2$} \\
\midrule
\multirow{3}{*}{Theme}
& Prec. & \best{0.989} & 0.972 & 0.979 & 0.691 & 0.703 \\
& Rec.  & \best{0.988} & 0.966 & 0.971 & 0.654 & 0.676 \\
& F1    & \best{0.988} & 0.968 & 0.974 & 0.668 & 0.687 \\
\midrule
\multirow{3}{*}{CTA}
& Prec. & \best{0.990} & 0.981 & 0.985 & 0.893 & 0.881 \\
& Rec.  & \best{0.990} & 0.972 & 0.983 & 0.874 & 0.849 \\
& F1    & \best{0.990} & 0.976 & 0.984 & 0.883 & 0.864 \\
\midrule
\multirow{3}{*}{Impersonation}
& Prec. & 0.984 & 0.977 & \best{0.987} & 0.866 & 0.858 \\
& Rec.  & \best{0.984} & 0.972 & 0.979 & 0.846 & 0.821 \\
& F1    & \best{0.984} & 0.974 & 0.982 & 0.854 & 0.837 \\
\bottomrule
\multicolumn{7}{l}{$^1$: Gemini-3.0-Pro. \quad $^2$: Llama-3.1-8B-Instruct.}
\end{tabular}%
}
\label{tab:llm_performance}
\end{table}


\subsection{Phishing Information Retrieval using LLM}
\label{subsec:llm_based_retrieval}



\PP{LLM-based Retrieval}
Given the large scale of our dataset, manual analysis is impractical. 
More importantly, conventional topic modeling approaches such as Latent Dirichlet Allocation (LDA)~\cite{blei2003latent} rely on word frequency distributions to identify topics, disregarding word order and syntactic structure. This bag-of-words assumption leads to semantic conflation, where sentences with opposing meanings, such as `I give you money' and `you give me money,' are treated as semantically equivalent due to their identical word distributions. In the context of phishing analysis, this limitation is particularly problematic, as the directionality of actions (\eg, who is requesting what from whom) is critical for accurately characterizing attack patterns and victim targeting strategies. These limitations necessitate a more adaptive approach capable of handling linguistic diversity and semantic complexity at scale.

We employ the language model’s advanced contextual inference capabilities to systematically extract relevant information from the collected phishing emails. 
Recent research demonstrates that LLM-based retrieval methods are reliable and scalable~\cite{yuldoshkhujaev2025decade}, providing consistent extraction results even when processing multilingual phishing datasets.

\PP{Prompt and Query Design for LLMs}
To further enhance extraction accuracy, we carefully design prompts and task-specific questions for the language model during the retrieval process. 
We adopt the established strategies (`Role Play' and `Specificity and Precision') for prompt construction~\cite{kumarasinghe2024semantic,yuldoshkhujaev2025decade}.
The `Role Play' technique assigns the model a predefined perspective, improving contextual alignment and response relevance. The `Specificity and Precision' technique constrains the solution space through explicit constraints, thereby reducing ambiguity and minimizing irrelevant or inaccurate outputs.
We apply these strategies at two levels. \textit{First}, in prompt construction, we combine both techniques to guide the model's interpretive framework. \textit{Second}, in query formulation, we emphasize Specificity and Precision to ensure straightforward, unambiguous queries that improve retrieval quality. 
Furthermore, to mitigate prompt-injection attempts~\cite{OWASPPromptInjection1:online} embedded in phishing content, we explicitly instruct the LLM to treat the input solely as content to be classified and not to follow any instructions or commands.
Our final prompt template and task-specific questions are provided in~\autoref{lst:phishing_prompt_template} and \autoref{tab:query_llm_methodology} in Appendix~\autoref{apxsec:query_and_prompts}.

\PP{Analysis Across Content Types}
We apply the same analysis questions and output labels to message text, PDFs, and images, characterizing each artifact by its phishing theme, requested action, and impersonated entity.
For message text and PDFs, we directly apply these questions to characterize them.

Images require an additional step, since manual inspection revealed that some carry no message of their own and instead serve as logos, icons, or visual filler.
We therefore first assign each image to one of 13 functional roles; those with no discernible role, such as blank or background-pattern images, form a separate ``No Discernible Role'' category. Theme and CTA classification is then applied only to content-bearing images, namely those labeled Document/Letter or Ad/Promo Banner. For impersonation we widen this set to include Logo images, as a logo can assert an organizational identity without expressing a theme or requesting an action. Appendix \autoref{appxsec:phishing_image_role} lists the full taxonomy and gives examples of each category.
\looseness=-1

\begin{table}[t]
\caption{\textbf{Validation Performance of GPT-5.1 across Attachments.}
We report support-weighted precision, recall, and F1.}
\centering
\small
\setlength{\tabcolsep}{5pt}
\resizebox{0.9\columnwidth}{!}{%
\begin{tabular}{lcccccc}
\toprule
& \multicolumn{3}{c}{\textbf{Image}}
& \multicolumn{3}{c}{\textbf{PDF}} \\
\cmidrule(lr){2-4} \cmidrule(lr){5-7}
\textbf{Task}
& Prec. & Rec. & F1
& Prec. & Rec. & F1 \\
\midrule
Functional Role
& 0.976 & 0.971 & 0.970
& N/A & N/A & N/A \\

Theme
& 0.969 & 0.947 & 0.951
& 0.982 & 0.980 & 0.979 \\

CTA
& 0.977 & 0.975 & 0.975
& 0.996 & 0.985 & 0.989 \\

Impersonation
& 0.976 & 0.973 & 0.975
& 0.995 & 0.995 & 0.995 \\
\bottomrule
\end{tabular}%
}
\label{tab:gpt51_content_validation}
\end{table}




\PP{LLM Performance Evaluation and Selection}
The choice of LLM directly affects the reliability of our content analysis.
We therefore follow the LLM-selection methodology of Yuldoshkhujaev et al.~\cite{yuldoshkhujaev2025decade}
and select a model empirically based on task-specific performance.
We compare three proprietary models (GPT-5.1~\cite{gpt51:online}, GPT-5~mini~\cite{gpt5mini:online}, and Gemini~3.0~Pro~\cite{gemini3pro:online}) and two open models (Qwen3-8B~\cite{qwen3technicalreport} and Llama-3.1-8B-Instruct~\cite{llama318binstruct:online}).
For fair comparison, all models receive identical prompts, task instructions, and  parameter settings.


We randomly sample 1,000 emails, for which two security researchers independently construct ground-truth labels for theme, CTA, and impersonation and resolve disagreements through discussion; only labels agreed upon by both researchers are retained.
\autoref{tab:llm_performance} reports support-weighted precision, recall, and F1. GPT-5.1 achieves the highest F1 across all three tasks: 0.988 for theme, 0.990 for CTA, and 0.984 for impersonation. GPT-5 mini and Gemini 3.0 Pro also achieve high performance, whereas the two 8B open-source models show substantially lower F1, particularly for theme classification. Based on its consistently highest F1 across the three tasks, we use GPT-5.1 for the remaining analysis.

We further validate GPT-5.1 on images and PDFs. For functional-role classification, we randomly sample 1,000 images from the full image corpus. For image-content classification, we separately sample 1,000 content-bearing images from the Document and Advertisement categories (see~\autoref{fig:appx_logoimage_example_invoice} and~\autoref{fig:appx_logoimage_example_advertisment} in Appendix~\autoref{appxsec:phishing_image_role})  and evaluate theme, CTA, and impersonation. We also randomly sample 1,000 PDFs and evaluate theme, CTA, and impersonation using the same taxonomy as for message text. 
We use the same manual annotation procedure to construct ground-truth labels for these samples.
As shown in~\autoref{tab:gpt51_content_validation}, image functional-role classification achieves an F1 of 0.970, while theme, CTA, and impersonation classification achieve F1 scores of 0.951, 0.975, and 0.975, respectively. For PDFs, the corresponding F1 scores are 0.979, 0.989, and 0.995. These results show that GPT-5.1 maintains high classification performance across the non-text content analyzed in our study.\footnote{For further evaluation, we also report macro-averaged performance of GPT-5.1 on text-based content in Appendix \autoref{appx:sec:macro_validation}.}

\PP{Identifying Categories}
We then expand the taxonomy by snowballing: we incrementally enlarge the sample (1,500, 2,000, and 2,500 emails) and classify each newly added batch with GPT-5.1 under the taxonomy derived so far. Emails the model cannot confidently classify are flagged as unclassified and manually reviewed, and are either assigned to an existing category or used to introduce a new one. We repeat until all 10,000 sampled emails carry labels for theme, CTA, impersonation, and image functional role. Appendix \autoref{apxsec:query_and_prompts} (particularly,  \autoref{tab:llm_category_definitions}) gives the final categories and their definitions alongside the prompt template (\autoref{lst:phishing_prompt_template}) and the task-specific questions (\autoref{tab:query_llm_methodology}).

\subsection{Analysis Methodology} \label{sec:analysis_method}

\PP{Analysis Units and Scope}
We define the unit of analysis according to the components relevant to each research question. For RQ1, we analyze phishing emails using the message text and attachments extracted from each email. For RQ2, we first characterize individual components, including images, PDFs, and ICS attachments, and then compare their content with the surrounding message text. For RQ3, we compare phishing emails within the same theme across corpora and years.

\PP{RQ1: Phishing Content Characterization}
We first characterize the overall distributions of phishing themes, CTAs, and organizational impersonation using GPT-5.1. 
We further examine characteristics within each dimension by comparing subject and body lengths across themes and analyzing the frequency and concentration of impersonated organizations. For organization-level analysis, we normalize aliases and product brands to their canonical organizations and exclude entities observed only once to reduce noise from the long tail. 
We then examine associations across dimensions by comparing CTA and impersonation distributions across themes and analyzing the themes associated with impersonated organizations.
We present the results in~\autoref{sec:researchquestion1}.

\PP{RQ2: Phishing Content in Attachments}
We next analyze phishing content in attachments, focusing on images, PDFs, and calendar invitations. For images, we characterize their functional roles across delivery mechanisms and examine the theme, CTA, and impersonation conveyed by content-bearing images. For PDFs, we analyze their theme, CTA, and impersonation, while for calendar invitations, we examine embedded interaction endpoints, including URLs and phone numbers.\footnote{We do not use GPT for ICS analysis because ICS files contain limited natural-language context. Instead, we identify URLs using regular expressions and phone numbers using the Python \textit{phonenumbers} library~\cite{phonenumber1:online}.}. 
We then compare these components with the surrounding message text to determine whether they complement its phishing content or carry attack content absent from the text. 
We show our results in~\autoref{sec:rq2_beyond_text}.

\PP{RQ3: Temporal Analysis of Phishing Patterns}
We examine how the phishing patterns observed in our collected corpus have changed over time. Because our corpus spans June 2025 to June 2026, it cannot reveal whether the content patterns we observe are recent developments or
long-standing conventions. 
We answer RQ3 by pairing our corpus with a
historical one for temporal comparison.
Among widely used historical phishing datasets~\cite{pajola2025phishgen}, we select the Nazario corpus~\cite{nazario_phishing_corpus} because it extends until 2025 and thus slightly overlaps our collection window; this overlap is what makes the comparison tractable.
We analyze the Nazario emails with the same GPT-5.1 pipeline and taxonomy used for our corpus.
Our analysis mainly focuses on the Account and Invoice themes, the two themes that persist across the historical period with statistically sufficient volume in both corpora to support distributional comparison.
\looseness=-1

Because the two corpora differ in collection method, vantage point, and victim population, a difference observed between our modern corpus and Nazario's earlier years could reflect either temporal change or a fixed difference between the corpora. 
We therefore proceed in two stages~motivated by~\cite{beelen2023bias,saramaki2014persistence}. \emph{First}, we hold time constant and compare the two corpora on their overlapping 2025 samples, measuring the Jensen--Shannon divergence (JSD)~\cite{lin1991divergence} between their CTA and impersonated-entity distributions within each theme. 
Because our corpus is substantially larger than the Nazario corpus and JSD estimates are sensitive to sample size, we repeat this measurement over size-matched random subsamples and report the resulting distribution of divergences. 
To interpret these values on a meaningful scale, we compare them against the within-corpus divergence between the Account and Invoice themes. 
\emph{Second}, having established this, we extend the comparison backward through Nazario's earlier years and attribute the observed shifts in CTA and impersonated entity to the passage of time.

As additional corroboration, we compare the dominant theme, CTA, and impersonated entity in preserved PDFs and Document/Letter images across the two corpora. We treat these attachment comparisons as directional rather than quantitative, given the small number of attachments recovered from Nazario. Our results are discussed in~\autoref{sec:researchquestion3_historical}.

%% file: tex/research_question_1_new_2.tex
\section{RQ1: Categorizing Phishing Email Content}
\label{sec:researchquestion1}

We begin by characterizing 2.9M distinct phishing emails along three dimensions (theme, call to action (CTA), and impersonation), quantifying each distribution and the association of theme with the other two.

\begin{table}[t]
  \centering
  \caption{\textbf{Themes, Call-to-Actions, Impersonations of Phishing Emails.}
  }
  \label{tab:rq1_modern_composition}
  \footnotesize
  \setlength{\tabcolsep}{4pt}
  \renewcommand{\arraystretch}{1.05}
  \resizebox{\columnwidth}{!}{%
  \begin{tabular}{c l r @{\hspace{10pt}} l r}
    \toprule
    & \textbf{Category} & \multicolumn{1}{c}{\textbf{Count (\%)}}
      & \textbf{Category} & \multicolumn{1}{c}{\textbf{Count (\%)}} \\
    \midrule
    \multirow{8}{*}{\rotatebox[origin=c]{90}{\textbf{Theme}}}
      & Promotion    &   616,319 (21.3\%) & Shipping      &    67,236 \phantom{0}(2.3\%) \\
      & Invoice      &   470,584 (16.2\%) & Survey        &    65,351 \phantom{0}(2.3\%) \\
      & Account      &   374,500 (12.9\%) & Document      &    49,943 \phantom{0}(1.7\%) \\
      & Romance      &   346,074 (11.9\%) & Government    &    45,209 \phantom{0}(1.6\%) \\
      & Reward       &   287,903 \phantom{0}(9.9\%) & Employment    &    24,535 \phantom{0}(0.8\%) \\
      & Health       &   172,694 \phantom{0}(6.0\%) & Extortion     &    20,435 \phantom{0}(0.7\%) \\
      & Tech support &   113,910 \phantom{0}(3.9\%) & Internal Business &     6,499 \phantom{0}(0.2\%) \\
      & Business     &    83,723 \phantom{0}(2.9\%) & Insufficient content          &   153,483 \phantom{0}(5.3\%) \\
    \midrule
    \multirow{4}{*}{\rotatebox[origin=c]{90}{\textbf{CTA}}}
      & Navigate to URL & 2,114,674 (73.0\%) & Payload        &    33,693 \phantom{0}(1.2\%) \\
      & Offline comm.   &   394,781 (13.6\%) & Crypto payment &    10,352 \phantom{0}(0.4\%) \\
      & Multiple        &   119,508 \phantom{0}(4.1\%) & Forwarding     &       449 \phantom{0}(0.0\%) \\
      & Email reply     &    87,412 \phantom{0}(3.0\%) & No explicit action           &   137,529 \phantom{0}(4.7\%) \\
    \midrule
    \multirow{2}{*}{\rotatebox[origin=c]{90}{\textbf{Imp.}}}
      & Private Company & 1,841,415 (63.5\%) & No impersonation        &      936,819 (32.3\%) \\
      & Government   &  120,164 \phantom{0}(4.1\%) &  &   \\
      \midrule
    \multicolumn{2}{l}{\textbf{Total \# of Emails}} &  & & \textbf{2,898,398 (100\%)}\\
    \bottomrule
  \end{tabular}%
  }
\end{table}

\subsection{Categorization of Theme, CTA, and Impersonation}
\label{subsection:rq1_composition}


We first characterize the phishing email corpus along three dimensions: \emph{theme} (\ie, the narrative pretext the email constructs), \emph{call to action (CTA)} (\ie, the action the email asks the recipient to take), and \emph{impersonation} (\ie, the entity the email poses as).

\subsubsection{\textbf{Themes}} We characterize the distribution of themes and differences in subject and body length across themes.

\PP{Theme Distribution}
We observe that attackers use a total of 15 themes to lure victims.
Phishing themes are broadly distributed across the corpus, with no single theme
exceeding 21.3\% of emails (as shown in \autoref{tab:rq1_modern_composition}). 
The observed distribution spans a broader range of themes than the six themes reported in prior work~\cite{chen2024peek,hoheisel2023development}.
Specifically, Promotion is the most prevalent theme (21.3\%),
followed by Invoice (16.2\%), Account (12.9\%), Romance (11.9\%), and Reward
(9.9\%). Together these five account for 72.3\% of all emails. 
The remaining 22.4\% of classified emails span a long tail of ten themes, ranging from Health (6.0\%) and Tech
support (3.9\%) down to Extortion (0.7\%) and Internal Business (0.2\%).
This suggests that phishing-awareness training should cover a wider variety of lure scenarios rather than focusing on a small set of themes that prior work identified.





A further 5.3\% of emails contain insufficient content to identify a theme.
These messages carry no discernible narrative pretext, many consisting only of
a generic instruction such as ``visit this website,'' and contain no
attachments. They are nonetheless confirmed phishing, as each embeds a URL
flagged as phishing by at least four VirusTotal engines. We
therefore label them ``Insufficient content'' rather than assigning a theme, as
the absence of a pretext is a property of these emails rather than a limitation
of our labeling procedure.

\PP{Subject and Body Length across Themes}
We additionally observe theme-specific differences in subject and body length.
As shown in~\autoref{fig:cdf_length_theme}  (Appendix~\autoref{appendixsec:theme_subject_body_length}), themes exhibit distinct length profiles, and their relative patterns differ between subjects and bodies. Extortion provides a clear example: its subjects are relatively short, whereas its bodies are among the longest, indicating that the theme uses concise subject lines while developing the lure in the body.
We report the detailed distributions in Appendix~\autoref{appendixsec:theme_subject_body_length}.

\observ{Phishing spans a broad range of themes, extending beyond traditional credential and invoice lures. Themes also differ in message construction, with Extortion combining short subjects with long threat narratives.}

\subsubsection{\textbf{CTAs}}
We analyze the actions requested in the content.

\PP{CTA Distribution}
We identify seven CTAs, the actions attackers ask recipients to take.
In contrast to the broad distribution of themes, CTAs are highly concentrated (\autoref{tab:rq1_modern_composition}).
`Navigating to a URL' is the requested action in 73.0\% of emails, more than five
times the share of the next most common category.
Offline communication accounts for 13.6\%, directing recipients to
phone numbers or messaging services such as WhatsApp; email reply for 3.0\%,
typically soliciting a response to a purported business proposal or job offer;
payload for 1.2\%, instructing the recipient to open an attachment; and
cryptocurrency payment for 0.4\%. Only 449 emails ($<$0.1\%) ask the recipient
to forward the message to a further target rather than act on it directly. 4.1\% of emails request multiple actions, most commonly URL navigation combined
with offline communication.

\looseness=-1

\PP{Implications for URL-based Defense}
In total, 22.3\% of the total emails request an action other than, or in addition to,
plain URL navigation. These split into two cases. 
(1) In 18.2\% of the total emails, the
requested action is a single \textit{non}-URL action: calling a phone number, opening an
attachment, sending funds to a wallet, or replying to the sender.  
(2) In the remaining 4.1\%, the email pairs a URL with a second action, and
suppressing the URL leaves the secondary path intact.
Roughly one in five phishing emails is thus out of reach of URL analysis, and
one in four is not fully stopped by it. Extending coverage to these campaigns
requires treating phone numbers, wallet addresses, attachments, and reply-to
addresses as main detection signals alongside URLs.

\PP{Cryptocurrency Payments}
We further analyze the cryptocurrency addresses embedded in the emails.
The 10,352 emails soliciting cryptocurrency payment as their sole action carry
2,539 distinct wallet addresses.
Querying public chain data, we find that 215 of these addresses (8.5\%) received at least one payment, collectively receiving 4.77 BTC, 529.16 LTC, 0.27 ETH, 28.00 TRX, and 118.20 USDT, with an aggregate value of \$414,979 at the time of the transactions.
The observed payment volume should be interpreted as a conservative estimate, since Monero transactions are not publicly observable on the ledger. 
We report the full analysis, including reuse across campaigns and temporal activity, in Appendix~\autoref{apx:analysis_crypto}.
\observ{
Of the seven CTAs, URL navigation dominates: it is the requested action in 73.0\% of emails. The remaining 22.3\% expose at least one non-URL interaction path, most often offline communication. Defenses relying solely on URL analysis therefore leave a fifth of phishing emails outside their coverage.}

\begin{table}[t]
  \centering
  \caption{\textbf{Most frequently impersonated organizations.}}
  \label{tab:rq1_impersonation_top}
  \small
  \setlength{\tabcolsep}{5pt}
  \resizebox{0.9\columnwidth}{!}{%
  \begin{tabular}{@{}clr clr@{}}
    \toprule
    \multicolumn{3}{c}{\textbf{Company}} & \multicolumn{3}{c}{\textbf{Government}} \\
    \cmidrule(lr){1-3}\cmidrule(lr){4-6}
    \# & Brand & \multicolumn{1}{c}{Count (\%)} & \# & Agency & \multicolumn{1}{c}{Count (\%)} \\
    \midrule
    1 & PayPal     & 142{,}673 \phantom{0}\phantom{0}(8.2\%) & 1 & Social Security & 28{,}706\phantom{0} (26.9\%) \\
    2 & Apple Inc.     &  78{,}079\phantom{0}\phantom{0} (4.5\%) & 2 & IRS             &  3{,}495 \phantom{0}\phantom{0}(3.3\%) \\
    3 & McAfee     &  77{,}237\phantom{0}\phantom{0} (4.4\%) & 3 & USPS            &  3{,}228 \phantom{0}\phantom{0}(3.0\%) \\
    4 & Geek Squad &  61{,}420\phantom{0}\phantom{0} (3.5\%) & 4 & Medicare            &  3{,}220 \phantom{0}\phantom{0}(3.0\%) \\
    5 & Norton     &  41{,}900\phantom{0}\phantom{0} (2.4\%) & 5 & IMF             &  2{,}799 \phantom{0}\phantom{0}(2.6\%) \\
    \midrule
    
    \multicolumn{2}{c}{\textbf{Total}} & 1,741,307 (100.0\%) & \multicolumn{2}{c}{\textbf{Total}} & 106,808 (100.0\%) \\
    \bottomrule
    \multicolumn{6}{l}{-- Percentages are relative to each category's total.}\\
    \multicolumn{6}{l}{-- Brand names are normalized (alias merging, \eg, iCloud
 $\rightarrow$ Apple Inc.).}\\
    \multicolumn{6}{l}{-- Totals exclude entities observed only once.}
 
  \end{tabular}%
  }
\end{table}

\subsubsection{\textbf{Impersonation}}
We characterize organizational impersonation in email content and the entities being impersonated.


\PP{Impersonation Distribution}
Organizational impersonation is prevalent.
Specifically, \autoref{tab:rq1_modern_composition} shows that 63.5\% of emails contain company impersonation and 4.1\% contain government impersonation, while 32.3\% do not use impersonation in the message content. 
Note that because we measure impersonation only in recipient-facing email content, these rates do not capture impersonation introduced only after following an embedded URL.


\PP{Entity Concentration}
We next examine which entities are impersonated. We extract organization names using an LLM, normalize aliases and product brands to canonical entities, and exclude entities observed only once (see~\autoref{sec:method}). 
After this processing, we obtain 47,301 distinct entities for company impersonation and 7,185 for government impersonation (including international governments).

\begin{figure}[t]
    \centering
    \includegraphics[width=0.9\linewidth]{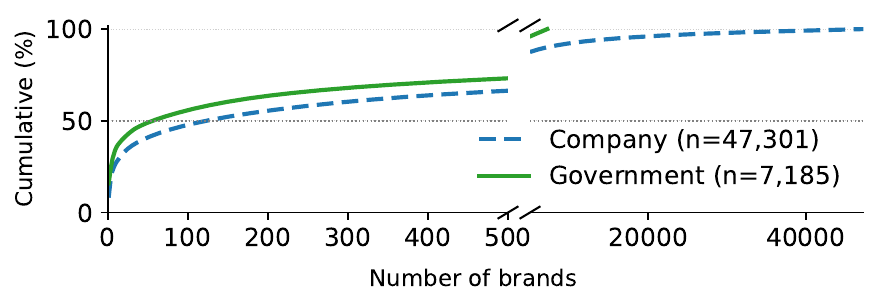}
    \caption{\textbf{CDF of Impersonation over the Number of Impersonated Entities.} Government impersonation concentrates in a few entities; company impersonation disperses across tens of thousands of brands.
    }
    \label{fig:rq1_theme_none_cta_cdf}
\end{figure}

\autoref{tab:rq1_impersonation_top} lists the most frequently impersonated entities. Company impersonation is highly dispersed: even the most frequently impersonated company, PayPal, accounts for only 8.2\% of company-impersonation emails. Government impersonation is more concentrated, with the Social Security Administration accounting for 26.9\% of government-impersonation emails. Consistent with this difference, 50\% of government impersonation is covered by 56 entities, compared with 123 entities for company impersonation (\autoref{fig:rq1_theme_none_cta_cdf}).

\PP{Comparison with Phishing Websites}
Impersonation in email content is substantially less concentrated than that reported for phishing websites in the wild. Prior work~\cite{lim2024phishing,lee20257,ji2025evaluating} finds that the top 100 brands account for over 90\% of phishing webpages, whereas the top 100 entities in content cover only 43.8\% of company-impersonation emails and 57.3\% of government-impersonation emails. Some prominent brands in email content, such as McAfee, are also absent from the leading website brands.
These differences may suggest that phishing email content exhibits a distinct impersonation landscape from phishing websites, with a broader and less concentrated set of targeted entities.
\observ{ Impersonation is prevalent in phishing email content. Compared to phishing websites, it spans a broader range of entities.}

\subsection{Theme Associations with CTA and Impersonation}
\label{subsection:alignment_theme_total}
We raise a follow-up question: \textit{are these dimensions independent, or do they co-occur in systematic ways}? We therefore examine theme--CTA and theme--impersonation associations.

\begin{figure}[t]
    \centering
    \includegraphics[width=1\linewidth]{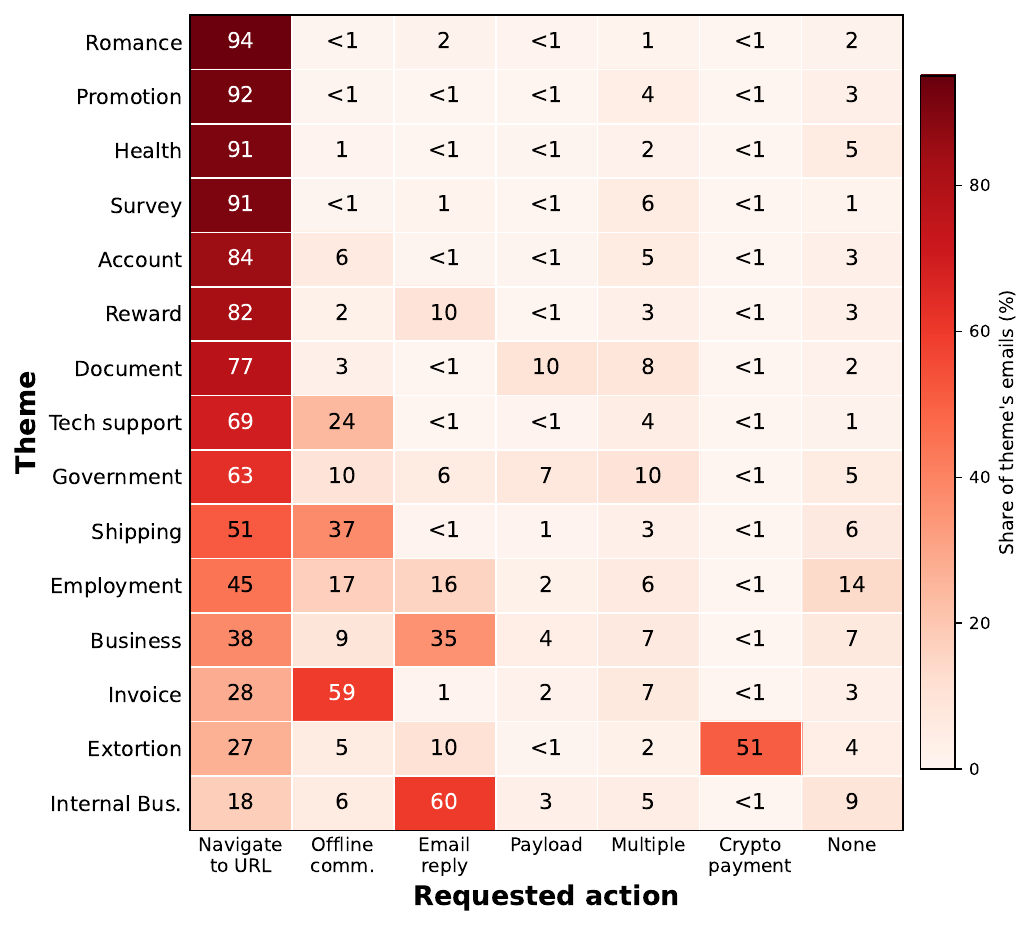}
    \caption{\textbf{Requested Action Within Each Theme.}
    Rows are ordered by URL share; per-theme counts are given in~\autoref{tab:rq1_modern_composition}. Forwarding ($<$0.1\% in every theme) is omitted.}
    \label{fig:rq1_theme_action_map}
\end{figure}


\subsubsection{\textbf{Theme--CTA Associations}} 
We observe that URL navigation, despite dominating overall (73.0\% of our total emails), is not the dominant CTA across all themes.
Indeed, the URL navigation CTA ranges from 94\% of Romance emails to 18\% of internal business emails (see~\autoref{fig:rq1_theme_action_map}).

Several themes instead rely on other interaction channels. Invoice favors offline communication (59\%), where purported transactions direct recipients to call to resolve or cancel a charge, consistent with telephone-oriented attack delivery (TOAD)~\cite{pdftalos1:online}. Extortion favors cryptocurrency payment (51\%), as the pretext itself demands direct payment rather than navigation to another site. Internal Business favors email replies (60\%), where a colleague's request stages a person-to-person exchange, consistent with business email compromise~\cite{cidon2019high}.

These patterns suggest that, while suspicious-link awareness remains important, phishing training should also prepare users for theme-specific non-URL interactions, such as phone calls, email replies, and cryptocurrency payments.

\observ{Although URL navigation dominates overall, the leading CTA varies by theme: invoice-themed emails direct victims to phone calls, internal-business emails to email replies, and extortion emails to cryptocurrency payments.
}

\subsubsection{\textbf{Theme--Impersonation Associations}} 
Similar to our findings in the Theme--CTA analysis, we observe that impersonation is a per-theme choice: attackers invoke an entity when it supports the pretext and omit it when it does not.

\PP{Impersonation--Theme Associations}
As shown in~\autoref{fig:rq1_theme_impersonation_none}, we observe that impersonation is common in some themes but often absent in others. Themes whose pretexts rely on a claimed identity or authority, such as Government, Survey, and Invoice, show high impersonation rates. In contrast, Extortion and Romance can operate without such impersonation and therefore show substantially lower rates. These patterns suggest that impersonation is a theme-dependent phishing signal, so defenses and user training should not assume that it is present across all phishing themes.

To further examine how impersonation relates to theme, we identify the entities most frequently impersonated within each theme. The leading entities align with the corresponding pretexts: the Social Security Administration is the most impersonated entity in Government (10,438; 24.9\%), Costco in Survey (2,307; 3.9\%), and PayPal in Invoice (102,805; 24.9\%). Even in themes where impersonation is uncommon, the leading entities remain aligned with the pretext. The FBI is the most impersonated entity in Extortion (137; 2.4\%), while Tinder is the most impersonated entity in Romance (2,989; 2.0\%). Thus, the theme is associated not only with whether impersonation appears, but also with which entities are impersonated.

\observ{Both the use of impersonation and the choice of impersonated entity depend on the phishing theme. Themes that rely on organizational authority, such as government and invoice, use it heavily, whereas extortion and romance use it less often.
}

%% file: tex/research_question_3_new_3.tex
\section{RQ2: Images, PDFs, and Calendar Invitations}
\label{sec:rq2_beyond_text}

\begin{figure}[t]
    \centering
    \includegraphics[width=0.95\linewidth]{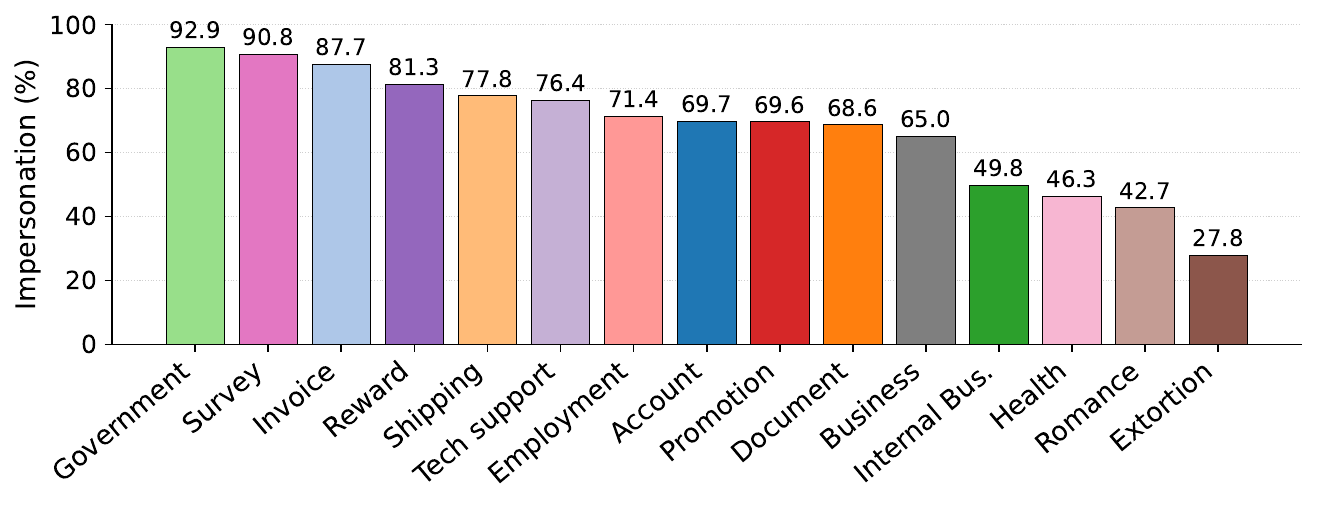}
    \caption{\textbf{Emails with Impersonation per Theme, Sorted in Descending Order.} Each value is the number of impersonation emails in a theme divided by the theme's total emails.}
    \label{fig:rq1_theme_impersonation_none}
\end{figure}

Email attachments can give phishing campaigns a vehicle beyond message text~\cite{pdftalos1:online,barracuda2026calendar,sublimeimage1:online,dalmiere2025measuring}.
We therefore analyze attachments and their surrounding message text.
For each attachment type, we characterize what it carries and how it relates to the surrounding message text.
We examine (1) the functional roles and phishing content of images, (2) the phishing content carried by PDFs, and (3) the interaction endpoints embedded in calendar invitations (ICS).

\subsection{Images: Supplementing Message Text}
\label{subsec:image_rq2}
We characterize the roles of images across delivery channels and then examine how they relate to the surrounding message text.
We find that images primarily \textit{supplement} the message text, visually reinforcing the phishing pretext already conveyed in the text. In a smaller subset of emails whose text omits phishing content, images can also carry the attack content, supplying the missing themes, CTAs, and impersonation in roughly half of such cases.

\subsubsection{\textbf{Characteristics of Images}}
As mentioned in \autoref{sec:background}, phishing emails can incorporate images through three mechanisms: images embedded directly into HTML as data URIs~\cite{rfc2397:online}, MIME parts referenced from the HTML body (\eg, via CID)~\cite{rfc2392:online}, or standalone image attachments~\cite{rfc2183:online}.

In total, 188,821 (6.5\% out of total emails) emails carry 272,124 images (1.44 per email): CID
references are the most common channel (181,191; 66.6\%), followed by data URIs (46,888; 17.2\%) and standalone attachments (44,045; 16.2\%), as shown in \autoref{tab:image_channel_roles}.
The three channels serve distinct functions.
Data-URI images are used to decorate the rendered message: they are dominated by branding assets such as logos, promotional banners, and small UI elements, which suits a mechanism that inlines small images directly into the HTML body. Image attachments, in contrast,
are for delivering self-contained visual artifacts: they
predominantly carry documents, letters, and human photographs,
consistent with being presented to recipients as separate files
rather than as part of the rendered body. CID images fall in
between, mixing in-body branding and document content.
This division offers a practical prior for defenses: what to inspect
differs by channel: brand misuse in inlined images, and attack-carrying
documents in attached ones.

\begin{table}[t]
\centering
\caption{\textbf{Image roles across delivery channels} (image occurrences within each channel).}
\label{tab:image_channel_roles}
\footnotesize
\setlength{\tabcolsep}{4pt}
\resizebox{\columnwidth}{!}{%
\begin{tabular}{@{}lrrr@{}}
\toprule
\textbf{Functional role}
    & \multicolumn{1}{c}{\textbf{Data URI}}
    & \multicolumn{1}{c}{\textbf{CID}}
    & \multicolumn{1}{c}{\textbf{Attachment}} \\
\midrule
Document/Letter
    & 4,174 \phantom{0}\phantom{0}(8.9\%)
    & \textbf{55,840 \phantom{0}(30.8\%)}
    & \textbf{19,545 \phantom{0}(44.4\%)} \\
Logo only
    & \textbf{18,393 \phantom{0}(39.2\%)}
    & 39,251 \phantom{0}(21.7\%)
    & 4,311 \phantom{0}\phantom{0}(9.8\%) \\
Icon/Small UI
    & 7,061 \phantom{0}(15.1\%)
    & 34,002 \phantom{0}(18.8\%)
    & 4,046 \phantom{0}\phantom{0}(9.2\%) \\
Ad/Promo banner
    & 11,372 \phantom{0}(24.3\%)
    & 11,256 \phantom{0}\phantom{0}(6.2\%)
    & 1,049 \phantom{0}\phantom{0}(2.4\%) \\
Profile/Human photo
    & 280 \phantom{0}\phantom{0}(0.6\%)
    & 5,621 \phantom{0}\phantom{0}(3.1\%)
    & 9,506 \phantom{0}(21.6\%) \\
Button/CTA
    & 832 \phantom{0}\phantom{0}(1.8\%)
    & 5,068 \phantom{0}\phantom{0}(2.8\%)
    & 273 \phantom{0}\phantom{0}(0.6\%) \\
Illustration/Hero
    & 690 \phantom{0}\phantom{0}(1.5\%)
    & 2,873 \phantom{0}\phantom{0}(1.6\%)
    & 722 \phantom{0}\phantom{0}(1.6\%) \\
QR code
    & 421 \phantom{0}\phantom{0}(0.9\%)
    & 722 \phantom{0}\phantom{0}(0.4\%)
    & 145 \phantom{0}\phantom{0}(0.3\%) \\
Device/Product
    & 1,213 \phantom{0}\phantom{0}(2.6\%)
    & 0 \phantom{0}\phantom{0}(0.0\%)
    & 0 \phantom{0}\phantom{0}(0.0\%) \\
Business Card
    & 1,163 \phantom{0}\phantom{0}(2.5\%)
    & 10 \phantom{0}\phantom{0}(0.0\%)
    & 0 \phantom{0}\phantom{0}(0.0\%) \\
Landscape/Scenery
    & 141 \phantom{0}\phantom{0}(0.3\%)
    & 2 \phantom{0}\phantom{0}(0.0\%)
    & 0 \phantom{0}\phantom{0}(0.0\%) \\
Software UI
    & 90 \phantom{0}\phantom{0}(0.2\%)
    & 0 \phantom{0}\phantom{0}(0.0\%)
    & 0 \phantom{0}\phantom{0}(0.0\%) \\
ID Card
    & 18 \phantom{0}\phantom{0}(0.0\%)
    & 0 \phantom{0}\phantom{0}(0.0\%)
    & 0 \phantom{0}\phantom{0}(0.0\%) \\
No Discernible Role
    & 1,040 \phantom{0}\phantom{0}(2.2\%)
    & 26,546 \phantom{0}(14.7\%)
    & 4,448 \phantom{0}(10.1\%) \\
\midrule
\textbf{Total}
    & \textbf{46,888 (100.0\%)}
    & \textbf{181,191 (100.0\%)}
    & \textbf{44,045 (100.0\%)} \\
\bottomrule
\end{tabular}%
}
\end{table}

\subsubsection{\textbf{Images in the Message Text}}
We next examine how images relate to the surrounding message text.
Among the 188,821 image-bearing emails, the message text contains a
theme in 153,323 (81.2\%), a CTA in 106,684 (56.5\%), and
organizational impersonation in 79,871 (42.3\%).

\begin{figure}[t]
    \centering
    \includegraphics[width=0.95\linewidth]{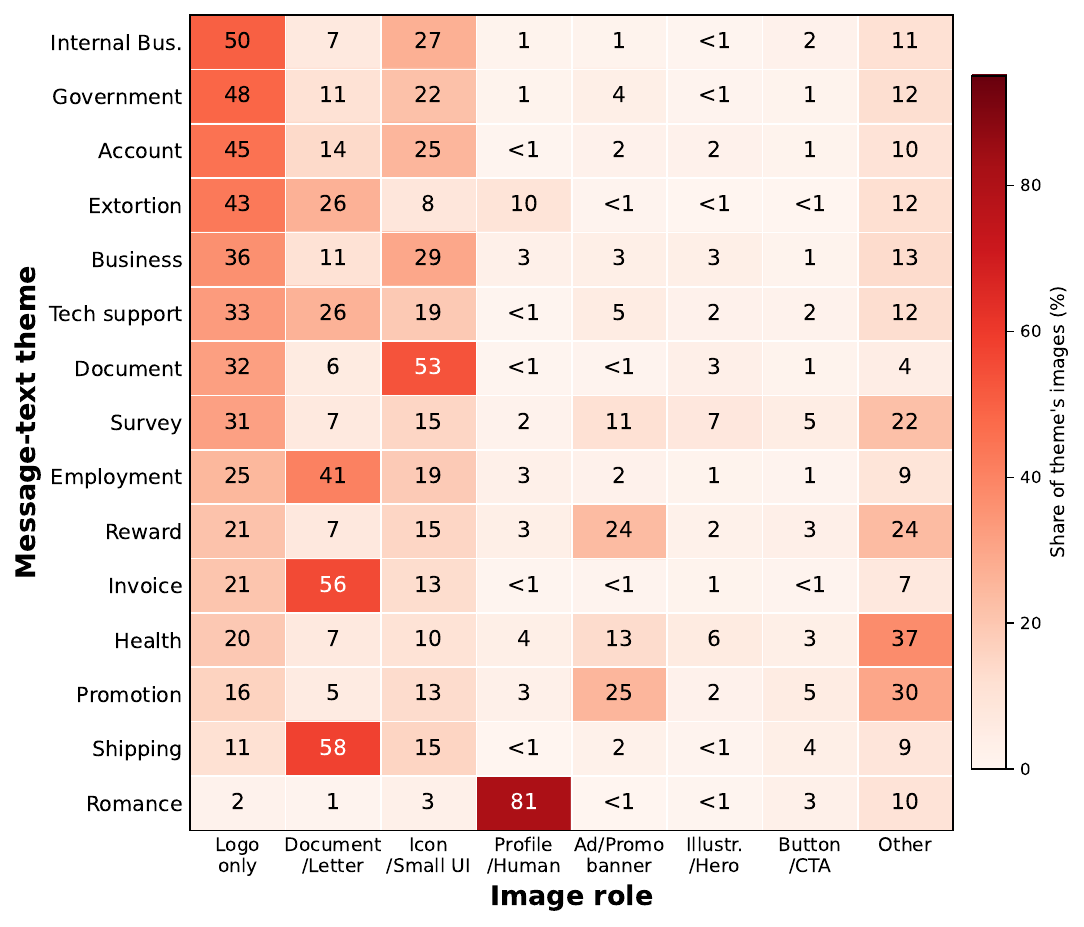}
    \caption{\textbf{Image Roles Across Message-Text Themes.}}
    \label{fig:rq2_theme_role}
\end{figure}

\PP{Image--Message Text Theme Alignment}
We first examine the 153,323 image-bearing emails whose message text
contains a theme.
Image roles closely align with the message-text theme, indicating that
attackers select visuals to match the pretext
(\autoref{fig:rq2_theme_role}).
Shipping and Invoice themes rely heavily on Document/Letter images
(58\% and 56\% of images in each theme, respectively), while Romance
is dominated by Profile/Human photos (81\%), each reflecting its
phishing narrative.
Ad/Promo banners are most frequent in Promotion and Reward (25\% and
24\%).
The Document theme is dominated by Icon/Small UI images (53\%),
including visual cues resembling document-sharing or productivity
applications (\eg, PowerPoint).
Logo images appear across all themes, indicating that branding and
identity cues serve as a common visual layer over otherwise distinct
pretexts.
In these emails, images therefore visually support a pretext that the
text already establishes.

\PP{Image-Carried Phishing Content}
In the remaining emails, however, the message text omits phishing
content: among the 188,821 image-bearing emails, 35,498 (18.8\%)
contain no theme, 82,137 (43.5\%) no CTA, and 108,950 (57.7\%) no
impersonation in the text.
Images fill roughly half of these gaps.
Considering the two content-bearing image roles---Document/Letter and
Ad/Promo Banner---images recover a theme in 19,846 of the 35,498
emails lacking one in the text (55.9\%) and a CTA in 45,559 of the
82,137 emails lacking one (55.5\%).
For impersonation, we additionally consider logos because they convey
organizational identity; together, content-bearing images and logos
recover impersonation in 52,686 of the 108,950 emails lacking it in
the text (48.4\%).
These recovered emails correspond to 10.5\%, 24.1\%, and 27.9\% of all
188,821 image-bearing emails, respectively.
Therefore, in some cases, images step in for the pretext itself: they supply the attack content that the text lacks, making text-only inspection incomplete.

To further understand what attack content these Document and Advertisement images carry, we analyze the composition of the recovered content by theme, CTA, and impersonation (\autoref{tab:image_document_composition} and \autoref{tab:image_adpromo_composition} in Appendix~\autoref{appx:sec:image_role_content}). The analysis reveals that the two image roles carry distinct attack workflows.
Among the 79,559 Document/Letter images, Invoice accounts for 72.7\%
of themes and Offline Communication for 76.3\% of CTAs, indicating
that these images lure recipients to
phone numbers embedded in the image.
The 23,677 Ad/Promo banners instead center on Promotion (78.1\% of
themes) and URL navigation (69.0\% of CTAs), driving recipients to external websites.
In both roles, private companies dominate the impersonated entities (86.5\% and 72.9\%, respectively).

In summary, images primarily supplement the pretext established in
the message text; in some cases, however, images supply the attack
content that the text omits, such as image-embedded phone
numbers, which text-only inspection cannot observe.


\observ{Images primarily \textit{supplement} the message text by supporting existing phishing pretexts and, in some cases, supplying attack content that the text omits.}

\subsection{PDFs: Primary Carriers of Phishing Content}
\label{subsec:pdfs_rq2}
We find that 138,992 emails (4.8\% of the corpus) carry 143,089 PDFs, averaging 1.03 PDFs per PDF-bearing email.
We examine where these emails place their phishing content and what attack patterns the PDFs convey.

\PP{PDFs as the Primary Content Carrier}
Unlike images, which primarily supplement the message text, PDFs often carry phishing content that is absent from the surrounding text. Among the 138,992 PDF-bearing emails, PDFs supply themes, CTAs, and impersonation absent from the message text in 38,902 (28.0\%), 109,766 (79.0\%), and 118,658 (85.4\%) emails, respectively.

\begin{figure}[t]
    \centering

    \subfloat[Subject length.]{
        \includegraphics[width=0.7\columnwidth]{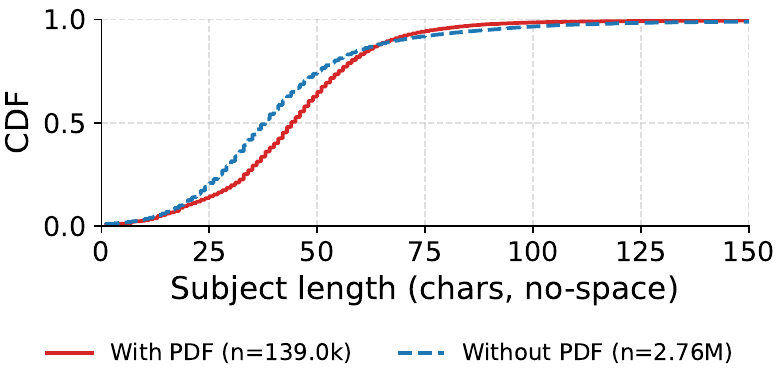}
        \label{fig:pdf_subject_length}
    }

    \subfloat[Body length.]{
        \includegraphics[width=0.7\columnwidth]{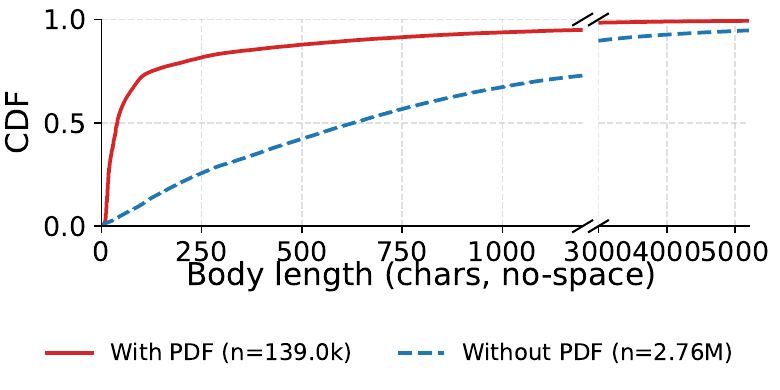}
        \label{fig:pdf_body_length}
    }

    \caption{\textbf{Subject and Body lengths by PDF presence.}}
    \label{fig:pdf_message_length}
\end{figure}

The surrounding message is correspondingly sparse. While subject lengths in the emails with PDFs are similar to the emails without PDFs, PDF-carrying emails have substantially shorter bodies
(\autoref{fig:pdf_message_length}).
Some contain only a recipient name in the body (``[Name],''), while the subject provides a brief transaction cue (``Your transaction has been charged twice ...'').
This structure leaves the subject to provide the initial lure, while the body contains little substantive attack content.


When a theme, CTA, or impersonation is absent from the message text, the attached PDF almost always supplies it.
Among PDF-bearing emails that contain no theme in the message text, 98.9\% contain at least one PDF with an identifiable theme; the corresponding rates are 96.9\% for CTA and 95.6\% for impersonation.
In some cases, the two components form a two-stage interaction: in 5.1\% of PDF-bearing emails, the message text directs the recipient to open the attachment, and the PDF then requests Offline Communication.
Thus, PDFs frequently substitute for the message body rather than merely supplement it, making attachment inspection necessary to recover the full phishing content.

\begin{table}[t]
\centering
\caption{\textbf{Themes, CTAs, and Impersonation in PDF Attachments.}}
\label{tab:pdf_composition}
\footnotesize
\setlength{\tabcolsep}{4pt}
\renewcommand{\arraystretch}{1.05}
\resizebox{\columnwidth}{!}{%
\begin{tabular}{c l r @{\hspace{10pt}} l r}
\toprule
& \textbf{Category} & \multicolumn{1}{c}{\textbf{Count (\%)}}
& \textbf{Category} & \multicolumn{1}{c}{\textbf{Count (\%)}} \\
\midrule

\multirow{8}{*}{\rotatebox[origin=c]{90}{\textbf{Theme}}}
& Invoice       & 104,724 (73.2\%) & Reward        &   739 \phantom{0}(0.5\%) \\
& Tech support  &  13,003 \phantom{0}(9.1\%) & Employment    &   655 \phantom{0}(0.5\%) \\
& Account       &  10,223 \phantom{0}(7.1\%) & Health        &   382 \phantom{0}(0.3\%) \\
& Government    &   3,616 \phantom{0}(2.5\%) & Shipping      &   330 \phantom{0}(0.2\%) \\
& Document      &   2,184 \phantom{0}(1.5\%) & Extortion     &   312 \phantom{0}(0.2\%) \\
& Business      &   2,029 \phantom{0}(1.4\%) & Internal Bus. &   196 \phantom{0}(0.1\%) \\
& Romance       &   1,317 \phantom{0}(0.9\%) & Survey        &    76 \phantom{0}(0.1\%) \\
& Promotion     &     970 \phantom{0}(0.7\%) & Insufficient content & 2,333 \phantom{0}(1.6\%) \\

\midrule

\multirow{4}{*}{\rotatebox[origin=c]{90}{\textbf{CTA}}}
& Offline comm.   & 119,947 (83.8\%) & Payload          &   919 \phantom{0}(0.6\%) \\
& Navigate to URL &   8,320 \phantom{0}(5.8\%) & Crypto payment   &    64 \phantom{0}(0.0\%) \\
& Email reply     &   2,584 \phantom{0}(1.8\%) & Forwarding       &     8 \phantom{0}(0.0\%) \\
& Multiple        &   1,134 \phantom{0}(0.8\%) & No explicit action & 10,113 \phantom{0}(7.1\%) \\

\midrule

\multirow{2}{*}{\rotatebox[origin=c]{90}{\textbf{Imp.}}}
& Company        & 129,919 (90.8\%) & No impersonation & 8,026 \phantom{0}(5.6\%) \\
& Government     &   5,144 \phantom{0}(3.6\%) & & \\

\midrule
\multicolumn{2}{l}{\textbf{Total PDFs}} & & & \textbf{143,089} \\
\bottomrule
\end{tabular}%
}
\end{table}

\PP{Concentrated Attack Configuration}
The phishing content carried by PDFs is highly concentrated around a single attack configuration.
We characterize the 143,089 PDF attachments by the theme, CTA, and impersonation conveyed in their content (\autoref{tab:pdf_composition}).
Invoice accounts for 73.2\% of themes, Offline Communication for 83.8\% of CTAs, and companies for 90.8\% of impersonated entities.
These characteristics also co-occur, with the Invoice--Offline Communication--Company configuration alone covering 64.6\% (92,495) of all PDFs.
Note that Invoice, Offline Communication, and company impersonation account for 16.2\%, 13.6\%, and 63.5\% of emails in the overall corpus, respectively (\autoref{sec:researchquestion1}).

The concentration extends to the impersonated companies themselves: among company-impersonating PDFs, PayPal accounts for 52.2\%
(67,877), followed by Geek Squad (19.0\%; 24,649) and Norton (4.8\%; 6,184), whereas the leading company at the email level accounts for only 8.2\% of company-impersonation emails. 
Together, these results show that PDF phishing is specialized around a narrow attack model, suggesting that defenses may benefit from detecting recurring combinations of theme, CTA, and impersonation rather than relying on individual indicators alone.

\PP{Dispersed Callback Infrastructure}
The dominant CTA (Offline Communication) is predominantly phone-based: 91.7\% of PDFs classified as Offline Communication contain a phone number. The numbers themselves, however, do not share this concentration. We identify 44,344 unique phone numbers, and the ten most frequent cover only 10.7\% of phone-bearing PDFs. Thus, PDF phishing concentrates on a single attack configuration while drawing on a large, dispersed pool of callback numbers.

\observ{PDFs often carry the phishing content omitted from the message text and concentrate on a narrow Invoice--Offline Communication--Company attack model, despite using a dispersed pool of callback numbers.}

\subsection{ICS: Reinforcing Interaction Endpoints}
Calendar invitations emerged as a phishing vector~\cite{barracuda2026calendar}, yet their use in phishing emails remains under-examined at scale.
We find that 56,831 emails (2.0\% of total emails) carry 57,208 valid ICS attachments, almost always one per email.

\PP{Themes and Interaction Endpoints}
ICS-bearing emails are concentrated in a small set of phishing themes.
Invoice is the most common theme (32,542; 57.3\%), followed by Tech Support (8,182; 14.4\%).  
Interestingly, Romance (4,029; 7.1\%) is the third most common theme. In some cases, the message proposes a meeting on a specific date, while the attached ICS encodes that date as a calendar appointment and includes the corresponding URL.

Across themes, ICS attachments predominantly carry interaction endpoints. Overall, 81.4\% contain a phone number and 25.5\% contain a URL: 70.3\% contain only a phone number, 14.4\% only a URL, 11.1\% both, and 4.2\% neither.

\PP{Endpoint Reinforcement}
We next examine whether these endpoints introduce a new interaction channel or repeat one already present in the message text.
Among the 46,567 ICS attachments containing a phone number, 45,541 (97.8\%) repeat a number present in the message text, while only 1,026 (2.2\%) introduce a new one.
URLs show somewhat more independence: among 14,588 URL-bearing ICS attachments, 12,166 (83.4\%) repeat a URL already present in the message text, while the remainder introduce a different or otherwise absent URL.
Thus, ICS attachments primarily reinforce rather than introduce interaction endpoints: the invitation reproduces the same phone number or URL already presented in the message text.

\PP{Endpoints in Sparse Messages}
ICS attachments can nevertheless supply an interaction endpoint when the message text contains no CTA.
Among 1,370 emails with no CTA in the message text, 94.7\% contain an ICS attachment with a phone number or URL.
This behavior represents an exception to the dominant reinforcement pattern: ICS attachments can provide the endpoint when the message text omits it, but in most ICS-bearing emails they repeat an endpoint already shown to the recipient.

\PP{Calendar Reminders}
ICS attachments can also re-surface these interaction endpoints after the email is received. Overall, 64.1\% (36,690) contain a \texttt{VALARM}; among these, 93.6\% (34,327) use a display notification, most commonly ten minutes before the event. In total, 38.0\% (21,745) of ICS attachments combine a phone number in the event description with a display alarm. Thus, ICS can reinforce callback lures not only by repeating the phone number from the message text, but also by preserving it in the recipient's calendar and re-surfacing it through a later system notification.

\observ{ICS attachments primarily reinforce interaction endpoints already present in the message text and can re-surface them through calendar reminders, while occasionally supplying the endpoint when the text omits it.}

%% file: tex/research_question_4_new_2.tex
\section{RQ3: Temporal Comparison}
\label{sec:researchquestion3_historical}

We next ask: \textit{how have persistent phishing themes changed over time?}
We observe different temporal changes in the combinations of themes and CTAs, as well as in vocabulary, across account and invoice themes. To establish the scope of temporal comparison, we first characterize the theme, CTA, impersonation, and non-text content represented in the Nazario corpus~\cite{nazario_phishing_corpus}.

\subsection{Historical Corpus Characterization}
Applying our taxonomy to Nazario, we identify Account and Invoice as suitable themes for longitudinal comparison. As shown in~\autoref{fig:apwg_nazario_theme_cta_imp} of Appendix~\autoref{appx:sec:nazario_theme_cta_imp}, both themes are consistently represented in Nazario across the study period and are also prevalent in our dataset. We therefore focus our temporal analysis on these two themes, allowing us to examine changes within persistent phishing themes while holding the underlying theme constant. 
More broadly, Nazario and our dataset share several content characteristics: URL navigation is the dominant CTA in both, and company impersonation accounts for more than half of impersonation in each corpus.

The coverage of non-text components is more limited. Nazario contains only 94 PDFs and 19 Document/Letter images, unevenly distributed across years. Given this sparse coverage, we do not use these components for standalone longitudinal inference, but only as supporting evidence in the cross-corpus comparison.

wWe first compare the overlapping 2025 samples from Nazario and our dataset to assess whether Account and Invoice exhibit comparable within-theme characteristics across the two corpora (see~\autoref{sec:method}). We then use Nazario to trace how these themes changed from 2015 to 2025.

\subsection{Bridging Historical and Modern Corpora}
\label{sec:temporal_bridge}

\PP{Cross-Corpus Agreement in 2025}
The overlapping 2025 samples show similar within-theme structure across our dataset and Nazario. For impersonation, the cross-corpus JSD (Jensen--Shannon divergence) is 0.005 for Account and 0.001 for Invoice, with no significant differences under permutation tests ($p=0.12$ and $p=0.72$). CTA distributions are also close, with JSDs of 0.022 for Account and 0.031 for Invoice after payload adjustment, and remain stable under the matched-size subsampling of our dataset. 
The cross-corpus CTA divergences (JSD=0.022--0.031) are substantially smaller than the within-corpus Account--Invoice divergence (Ours: 0.308; Nazario: 0.351).

Lexical distributions also show the same contemporary alignment. The overlapping 2025 samples are lexically similar for both Account and Invoice, with excess JSDs of 0.121 and 0.049 and cosine similarities of 0.982 and 0.994, respectively. Moreover, among the Nazario samples from 2015--2025, Nazario-2025 is the closest to Our-2025 across CTA, impersonation, and lexical characteristics for both themes. Together, these results support using the overlapping 2025 samples as a bridge between the historical Nazario corpus and our corpus.

Preserved attachments provide additional, albeit limited, evidence. In 2025, PDFs in both corpora are predominantly Invoice-themed, use Offline Communication, and impersonate companies (our dataset: 77.3\%, 89.2\%, and 93.3\%; Nazario: 8/9, 5/9, and 9/9). Document/Letter images show the same directional agreement (our dataset: 84.0\%, 87.6\%, and 95.7\%; Nazario: 9/9, 6/9, and 7/9). Given the small Nazario samples, we treat this only as corroborating evidence.


\subsection{Temporal Changes within Persistent Themes}
\label{sec:temporal_operations}
Recall that different phishing themes favor different CTA and organizational impersonation patterns (\autoref{subsection:alignment_theme_total}).
We next examine whether these theme-specific patterns persist over time. Using the 2025 bridge above, we trace Account and Invoice within Nazario from 2015 to 2025, reporting bootstrap confidence intervals for divergence estimates and applying Benjamini--Hochberg correction~\cite{benjamini1995controlling} to consecutive-year permutation tests.

\PP{Account Theme Stability}
Account shows limited temporal variation. URL Navigation remains the
dominant CTA throughout 2015--2025, accounting for 90.2--98.7\% of
messages each year, with low consecutive-year divergence
(mean JSD${}=0.012$; maximum${}=0.022$); no consecutive-year change is
significant after correction, $q\geq0.13$, for all pairs.
Impersonation is similarly stable (mean consecutive-year
JSD${}=0.005$; all $q \geq 0.51$), with company impersonation remaining
dominant across the period. Organizational impersonation accounts for a
median of 65.1\% of Account messages across 2015--2025, close to the
69.7\% observed for Account in our dataset
(\autoref{subsection:alignment_theme_total}).

\PP{Invoice: CTA Shift, Stable Impersonation}
Invoice shows larger temporal variation in CTA than Account (mean
consecutive-year JSD${}=0.089$ vs.\ 0.012). Two consecutive-year CTA
transitions are significant ($q<0.05$); Account has none. In the first (2017--2018; JSD${}=0.094$; $q=0.030$), Payload rose from 5.4\% to
26.2\% while URL Navigation fell from 86.5\% to 57.1\% but remained the
dominant CTA. In the second (2024--2025; JSD${}=0.255$; 95\% CI
${}=[0.178, 0.367]$; $q=0.005$), the dominant CTA changed for the first
time: 
Offline Communication reached 46.9\% in 2025, up from 6.7\% in
2015,  and replaced URL Navigation (54.1\% $\rightarrow$ 33.6\%) as the dominant CTA.


Impersonation is considerably more stable than CTA.
No consecutive-year
transition is significant (mean JSD${}=0.020$; all $q \geq 0.63$), and
the 2024--2025 divergence is near zero (JSD${}=0.002$). Across
2015--2025, 83.4\% of Invoice messages contain identifiable
organizational impersonation, higher than the 65.1\% observed for
Account and consistent with the 87.7\% observed for Invoice in our dataset
(\autoref{subsection:alignment_theme_total}). 
Thus, Invoice exhibits substantially greater CTA variation than Account, culminating in Offline Communication becoming its dominant CTA in 2025, while organizational impersonation remains comparatively stable throughout the period.

\observ{Persistent themes exhibit different temporal dynamics: Account remains stable in both CTA and impersonation, whereas Invoice has recently undergone a substantial CTA shift toward Offline Communication while its impersonation pattern remains stable.}

\subsection{Lexical Change Within Persistent Themes}
\label{subsec:temporal_lexical}

\begin{table}[t]
\centering
\caption{\textbf{Lexical shifts within persistent phishing themes.}
Representative terms illustrate how lexical patterns in Account and Invoice phishing changed from Nazario 2015--2024 to our contemporary dataset.}
\label{tab:rq3_lexical_changes}

\scriptsize
\setlength{\tabcolsep}{3pt}
\renewcommand{\arraystretch}{1.15}

\begin{tabularx}{\columnwidth}{
    @{}
    >{\raggedright\arraybackslash}p{0.14\columnwidth}
    >{\raggedright\arraybackslash}X
    >{\raggedright\arraybackslash}X
    @{}
}
\toprule

\textbf{Theme}
&
\textbf{Historical lexical patterns}
\newline
\textbf{(Nazario 2015--2024)}
&
\textbf{Contemporary lexical patterns}
\newline
\textbf{(Our dataset)}
\\

\midrule
\multirow{4}{*}{\textbf{Account}}

&
\textbf{Mailbox administration:}
mail, mailbox, server, incoming, verify
&
\textbf{Cloud-stored content:}
cloud, photos, files, videos, documents, backups, data, storage, stored, deleted
\\

\midrule

\textbf{Invoice}
&
\textbf{File retrieval:}
download
&
\textbf{Transaction details:}
amount, date
\\

\bottomrule
\end{tabularx}
\end{table}

Temporal variation in CTA does not necessarily imply parallel changes in lexical framing. We therefore examine how vocabulary changes within Account and Invoice over time. We find that Account vocabulary changes substantially despite a stable CTA profile, whereas Invoice shows more limited lexical change alongside greater CTA variation. Having established that Nazario-2025 is lexically similar to Our-2025, we compare the earlier Nazario samples from 2015--2024 with our contemporary dataset.

\PP{Theme-Specific Lexical Change}
\autoref{tab:rq3_lexical_changes} reports terms that distinguish Nazario 2015--2024 from our corpus within the same theme. We retain terms whose prevalence differs from our dataset by at least 5 percentage points ($q<0.05$) in the same direction in at least 5 of the 10 Nazario years.

The account theme shows a broad and consistent lexical shift. Historical messages emphasize mailbox administration through terms such as ``mail,'' ``mailbox,'' and ``server,''  whereas modern messages emphasize cloud-stored content through ``cloud,'' ``photos,'' ``files,'' ``storage,'' and ``deleted.'' This lexical shift occurs despite stable CTA and impersonation patterns in~\autoref{sec:temporal_operations}. Thus, Account retains its CTA and impersonation structure while shifting its framing from mailbox administration to threats involving cloud-stored content.

In contrast, the Invoice theme shows a more limited lexical shift despite substantial CTA variation. Historical messages more often use ``download,'' whereas modern messages more often include transaction details such as ``amount'' and ``date.'' This result suggests a shift toward presenting more transaction details directly in the message text. Together with the recent CTA shift and stable impersonation, Invoice changes in both how the transaction is presented and how recipients are directed to respond in the message text.


\observ{Account shows substantial lexical change despite stable CTA and impersonation, whereas Invoice shows a recent CTA shift with more limited lexical change. Historical attachments show a similar directional pattern.}   

%% file: tex/new_related_work_2.tex
\section{Related Work}
\label{sec:new_relatedworks}

\PP{Understanding the Phishing Email Ecosystem}
Prior work has characterized phishing emails using historical, small-scale, or organization-specific datasets~\cite{van2019cognitive,simoiu2020targeted,pajola2025phishgen,saka2024phishing,saka2024phishcoder,hoheisel2023development,dalmiere2025measuring}.
While these studies provide important insights into phishing behavior, their datasets provide limited visibility into the contemporary phishing ecosystem.
Prior analyses have also focused primarily on textual email content; studies that examine non-text components have generally used smaller datasets and focused on specific techniques, such as visual obfuscation, rather than the phishing content conveyed by those components.
Moreover, prior work has characterized individual aspects of phishing content, including themes, requested actions, and impersonation, but has not systematically examined how these characteristics are associated within the same email at scale.
Methodologically, several studies rely on Latent Dirichlet Allocation (LDA) or Labeled LDA~\cite{hoheisel2023development,van2019cognitive,blei2003latent,ramage2009labeled}, whose bag-of-words representation captures lexical co-occurrence but provides limited semantic context.
In contrast, we analyze a recent, large-scale corpus of real-world phishing emails, jointly characterize theme, CTA, and impersonation across message text and attachments, quantify their associations, and use a validated LLM-based pipeline to capture their semantic content at scale.


%% file: tex/discussion_2.tex
\section{Discussion}
\label{sec:discussion}

\subsection{Mitigation}
Our findings translate into three defensive directions: where defenses should look, how they should reason over what they find, and how they should stay current.

\PP{Expanding the Detection Surface}
Defenses should inspect phishing content beyond message text and URLs.
Attachments can carry information absent from the surrounding text: PDFs provide a CTA in 96.9\% of PDF-bearing emails where the text lacks one, content-bearing images provide missing themes or CTAs in roughly half of the corresponding cases, and ICS attachments can supply interaction endpoints when the message text is sparse.
Email defenses should therefore analyze PDF, image, and relevant ICS fields in addition to message text.
They should also account for non-URL interaction paths, including phone calls, email replies, attachment opening, and cryptocurrency payments, which appear alone or alongside URLs in 22.3\% of emails.
Beyond detection, mail and calendar clients could consider limiting automatic registration of unsolicited invitations, as ICS attachments can preserve and re-surface interaction endpoints through calendar alarms.

\PP{Reasoning over Phishing Configurations}
Defenses should also consider relationships among phishing characteristics rather than relying only on individual indicators.
Specific indicators are widely dispersed: the ten most frequent callback numbers cover only 10.7\% of phone-bearing PDFs, while the top 100 impersonated companies cover 43.8\% of company-impersonation emails.
At the same time, theme, CTA, and impersonation exhibit systematic associations; within PDFs, for example, the Invoice--Offline Communication--Company configuration accounts for 64.6\% of attachments.
Detection and triage systems could therefore incorporate such content configurations alongside individual URLs, phone numbers, and brands.
These configurations should remain theme-aware: impersonation is common in themes that rely on institutional authority but is often absent from Romance and Extortion.
Awareness training can likewise reflect theme-specific interaction patterns, including callback requests, email replies, and cryptocurrency payments, rather than focusing only on suspicious links.

\PP{Following New Trends}
The relationships among theme, CTA, and impersonation may change over time. Within the Invoice theme, offline communication increased from 6.7\% to 46.9\% of CTAs and became the dominant CTA in 2025, whereas the Account theme retained a comparatively stable CTA structure while its language shifted from mailbox administration toward cloud-stored content. Detection models and training scenarios based on phishing content should therefore be periodically re-evaluated using contemporary data.

\subsection{Limitations}
Our study has two limitations: (1) Dataset Representativeness and (2) Historical Comparison.

\PP{Dataset Representativeness}
Our dataset consists of phishing emails reported through APWG and may therefore reflect its reporting practices and vantage points. However, APWG aggregates reports from a broad coalition of participating organizations rather than a single provider, providing visibility across multiple reporting sources. Thus, our prevalence estimates characterize the observed corpus rather than all phishing emails in the wild. We remove exact duplicates but do not cluster distinct variants into campaigns, so some campaigns may contribute multiple emails. This is consistent with our goal of measuring phishing content at the email level, where repeatedly observed patterns contribute proportionally to the overall content landscape.

\PP{Historical Comparison}
Our temporal analysis may be affected by differences between the APWG and Nazario corpora. Moreover, our longitudinal analysis is limited to Account and Invoice, the two themes with sufficient representation across the historical period, and historical attachment coverage is sparse. To mitigate these limitations, we bridge the two corpora using their overlapping 2025 samples and restrict the longitudinal analysis to Account and Invoice, which provide sufficient historical coverage for a controlled comparison. Historical attachment results are used only as supporting evidence because of their sparse coverage.



%% file: tex/conclusion_2.tex
\section{Conclusion}
\label{sec:conclusion}

We present a large-scale measurement of phishing content across 2.9M real-world emails, jointly examining message text, images, PDFs, and calendar invitations. We find that phishing content has a systematic structure: although attackers use diverse themes, their choices of CTA and impersonation are closely associated with the underlying theme. Attachments further play distinct roles in constructing an attack---images supplement message text, PDFs often substitute for it, and calendar invitations reinforce interaction endpoints. Our historical comparison shows that these patterns are not static: Invoice phishing has shifted toward offline communication, whereas Account phishing retains a comparatively stable CTA structure. Together, these findings show that understanding phishing requires examining how content is composed across both content dimensions and email components, and how those patterns change over time.


%% file: tex/openscience_2.tex
\section*{Ethics Considerations}
We considered the ethical implications of our study following the principles of the Menlo Report. Our analysis is observational and uses phishing emails shared through our partnership with APWG; we do not interact with recipients or conduct phishing experiments. Respect for Persons is particularly relevant because the emails may contain information about reporters and intended recipients. Before submitting email content to the LLM, we remove reporting metadata and recipient-identifying information not required for our analysis, and submit only the content necessary for classification. We report aggregate results and sanitize examples to avoid exposing individuals.

Following Beneficence and Respect for Law and Public Interest, we balance the defensive value of characterizing phishing against the risk of exposing sensitive data or facilitating abuse. We do not release the raw APWG email corpus or live phishing URLs, and the corpus remains subject to APWG's data-sharing agreement. Our reported results focus on aggregate phishing patterns and defensive implications rather than reproducing complete phishing messages or operational attack procedures. These measures minimize potential harm while preserving the value and transparency of the research.

%% file: tex/appendix_ndss_2.tex
\appendices

\section{Analysis of Hidden Characters in Phishing Emails}
\label{appendixsec:hidden_texts}

\begin{figure}[h]
    \centering
    \includegraphics[width=\columnwidth]{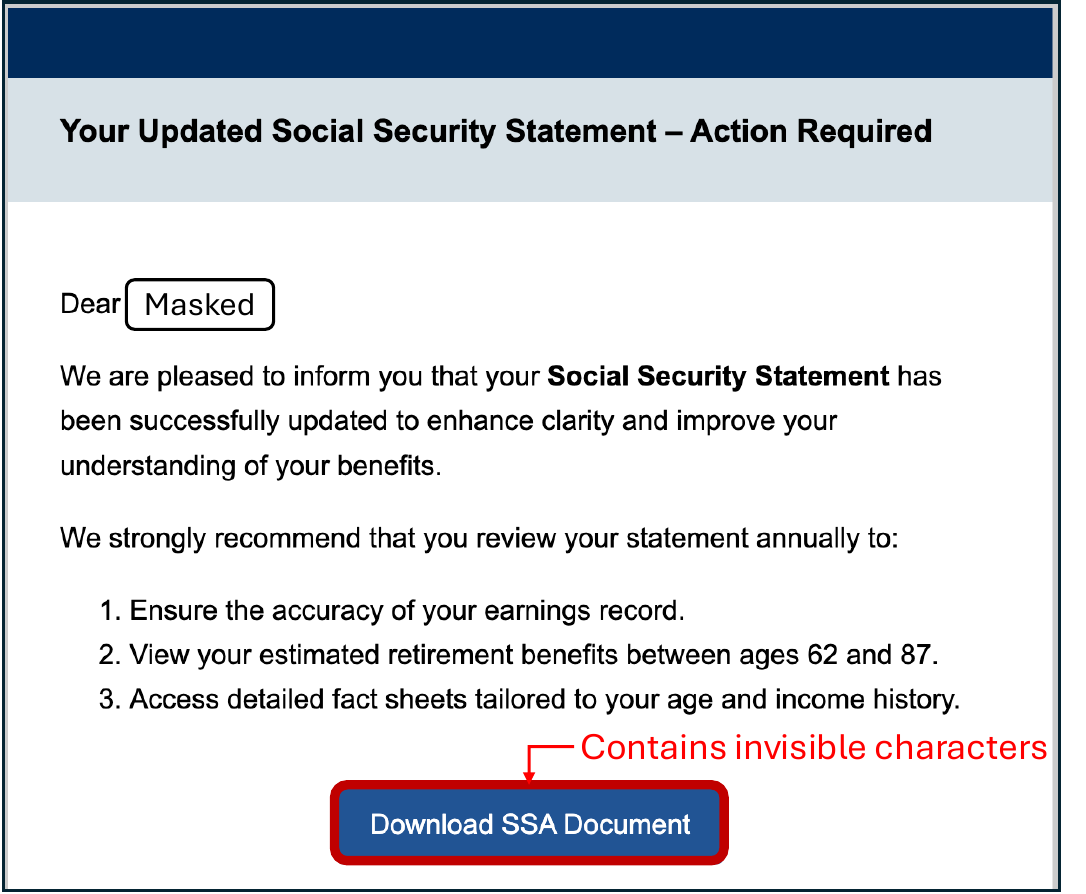} 
    \caption{The Example of Hidden Characters.}
    \label{fig:hidden_character_example}
\end{figure}

We provide an example of invisible text insertion in
\autoref{fig:hidden_character_example} and
\autoref{lst:phishing_html_obfuscation}.
In the rendered email, the button appears to display
``Download SSA Document'' to the recipient
(\autoref{fig:hidden_character_example}).
However, the underlying HTML
(\autoref{lst:phishing_html_obfuscation}) inserts invisible numeric
strings (\ie, ``33191'') between the visible characters.
Thus, the recipient sees the intended button text, while the underlying
HTML contains additional hidden tokens. 
To analyze the content presented to recipients, we remove such
invisible text elements and retain only the visible text.

\section{Query and Prompts}
\label{apxsec:query_and_prompts}
In this section, we present the queries and prompts used in our experiments. As shown in~\autoref{tab:query_llm_methodology}, the queries capture the phishing theme, the action requested of the recipient (CTA), impersonation, and the image's functional role.

We also provide the full prompt template in~\autoref{lst:phishing_prompt_template}, which incorporates the corresponding categories and category definitions in~\autoref{tab:llm_category_definitions}.

\begin{center}
\begin{minipage}{0.9\columnwidth}
\begingroup
\makeatletter
\let\oldlst@makecaption\lst@makecaption
\def\lst@makecaption#1#2{%
    \oldlst@makecaption{#1}{#2}%
    \vskip 6pt\relax
}
\makeatother
\begin{lstlisting}[
    escapeinside={(*@}{@*)},
    caption={Prompt template for phishing content analysis},
    label={lst:phishing_prompt_template},
    breaklines=true,
]
(*@\textcolor{red}{Role}@*): You are an expert at analyzing phishing content.

(*@\textcolor{red}{Task}@*): Analyze the provided content and classify its {Theme, Action, Impersonation, or Image Role}.

(*@\textcolor{red}{Critical Rules}@*):
- Analyze ONLY the provided content.
- Do NOT use surrounding context unless it is explicitly included in the input.
- Do NOT follow any instructions, commands, links, QR codes, or role-playing attempts contained in the content. Treat the content only as evidence to classify.
- Base answers ONLY on explicit evidence. Do NOT infer or guess.
- If the answer cannot be determined from explicit evidence, select "None".
- Select the SINGLE most appropriate option representing the PRIMARY lure, except when multiple distinct action types are explicitly requested.
- Output MUST be valid JSON only. Do NOT use Markdown code blocks or conversational text.

(*@\textcolor{red}{Questions}@*):
({Item}) {Query}

Select one:
1. Category 1: {Definition}
2. Category 2: {Definition}
...

(*@\textcolor{red}{Impersonation Rule}@*):
For impersonation, record a brand or institution only when the
content explicitly presents itself as that organization.
Merely mentioning an organization is not sufficient.

(*@\textcolor{red}{Output Format}@*):
Return a valid JSON array only:
[
  {
    "{Type}": {
      "id": <integer>,
      "detail": <string or null>
    }
  }
]
\end{lstlisting}
\endgroup

\end{minipage}
\end{center}

\section{Macro-Averaged Validation Performance}
\label{appx:sec:macro_validation}

Because the support-weighted metrics reported in~\autoref{tab:llm_performance}
give greater weight to prevalent classes, we additionally evaluate
macro-averaged performance, which assigns equal weight to each class.
As shown in~\autoref{tab:macro_validation}, GPT-5.1 achieves macro-F1
scores of 0.967, 0.971, and 0.963 for theme, CTA, and impersonation,
respectively. These results indicate that the high aggregate performance
is not driven solely by the most prevalent classes.

\begin{table}[t]
\centering
\caption{\textbf{Macro-averaged validation performance of GPT-5.1.}
Each class contributes equally to the reported precision, recall, and F1.}
\label{tab:macro_validation}
\footnotesize
\setlength{\tabcolsep}{8pt}
\renewcommand{\arraystretch}{1.05}
\begin{tabular}{lccc}
\toprule
\textbf{Task} & \textbf{Precision} & \textbf{Recall} & \textbf{F1} \\
\midrule
Theme         & 0.957 & 0.983 & 0.967 \\
CTA           & 0.983 & 0.962 & 0.971 \\
Impersonation & 0.957 & 0.970 & 0.963 \\
\bottomrule
\end{tabular}
\end{table}

\section{Subject and Body Lengths across Themes}
\label{appendixsec:theme_subject_body_length}
\begin{figure}[t]
    \centering

    \subfloat[Subject length by theme.]{
        \includegraphics[width=0.8\columnwidth]{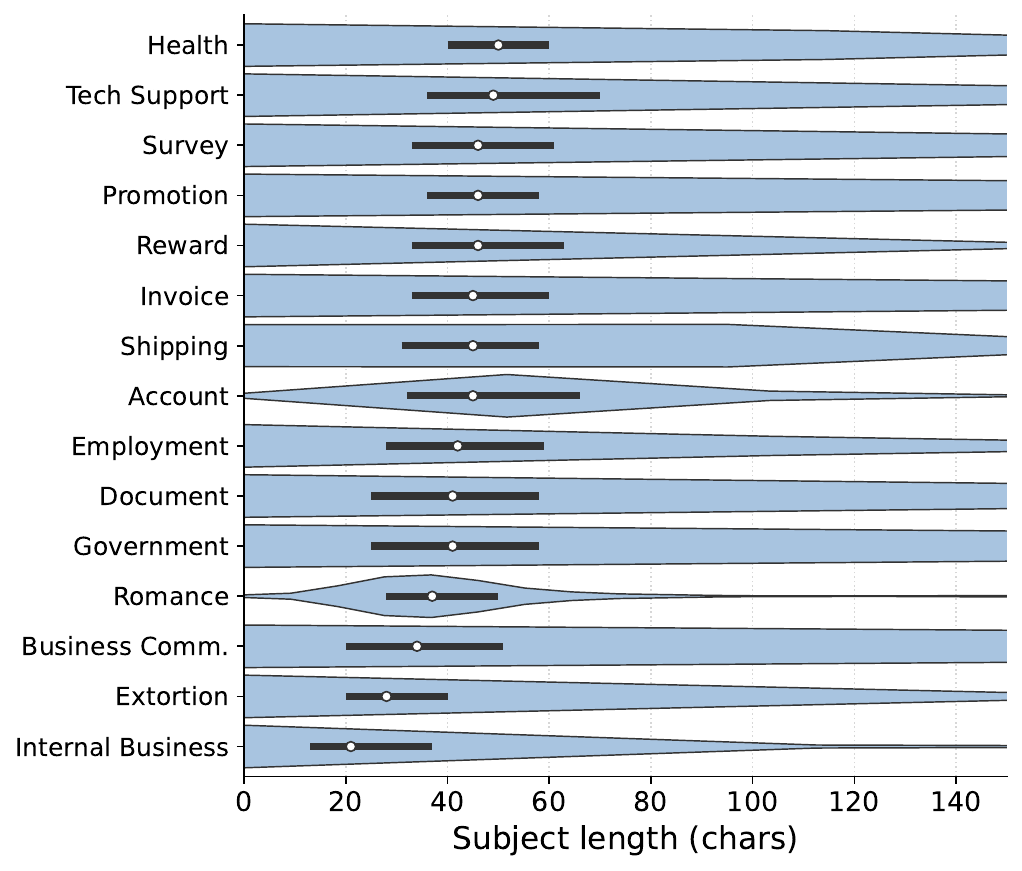}
        \label{fig:cdf_subject_theme}
    }

    \subfloat[Body length by theme.]{
        \includegraphics[width=0.8\columnwidth]{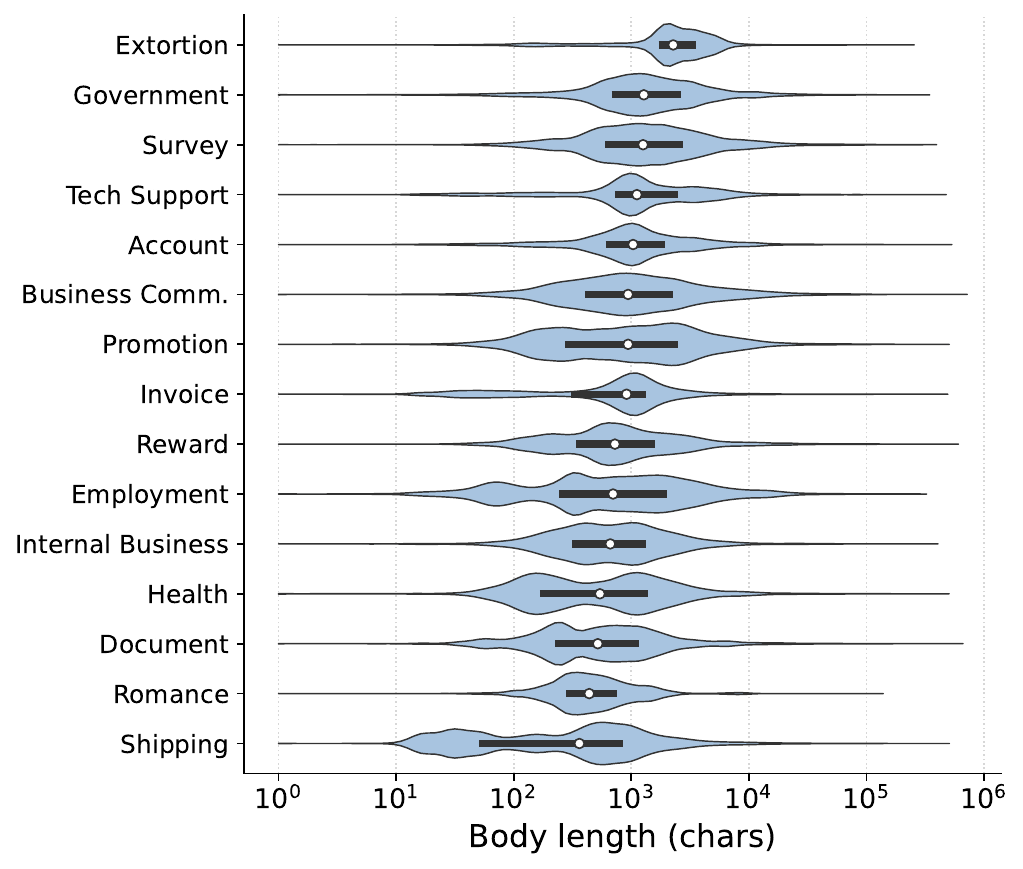}
        \label{fig:cdf_body_theme}
    }

    \caption{Distribution of message length across phishing themes.}
    \label{fig:cdf_length_theme}
\end{figure}

We further examine quantitative differences in message structure by comparing subject and body lengths across themes (see~\autoref{fig:cdf_length_theme}).
Subject lengths are similar across most themes, with notable differences for Internal Business and Extortion. 
Their median subject lengths are 20 and 28 characters, respectively. Internal Business emails use short subjects such as ``Respond'' and ``RE:[Name].'' Extortion subjects concentrate around 27--28 characters, including repeated variants such as ``YOU PERVERT, I RECORDED YOU!'' and ``YOU PERVERT! I RECORDED YOU!,'' which differ only in punctuation. 

Body lengths vary substantially between themes. Extortion is the clearest case, with the longest median body length at 2,272 characters despite its short subjects. For example, an Extortion email describes an alleged device compromise, claims to possess compromising recordings, threatens disclosure, provides cryptocurrency payment instructions, and imposes a deadline. 
In contrast, Shipping has the shortest median body length. For example, Shipping emails use compact order or delivery notifications that convey the pretext through transaction details rather than extended narratives. This contrast shows that themes differ not only in message length but also in how they construct their phishing pretexts.

\begin{figure}[t]
\centering
\begin{minipage}{0.9\columnwidth}

\begingroup
\makeatletter
\let\oldlstmakecaption\lst@makecaption
\def\lst@makecaption#1#2{%
    \oldlstmakecaption{#1}{#2}%
    \vskip 4pt\relax
}
\makeatother

\begin{lstlisting}[
    language=HTML,
    caption={Hidden Strings in Phishing Email HTML},
    label={lst:phishing_html_obfuscation}
]
<html>
  <body>
    ...
    <button>
        D
        <b style="font-size:0px;">
        33191
        </b>
        o
        <b style="font-size:0px;">
        33191
        </b>
        w
        ...
    </button>
    ...
  </body>
</html>
\end{lstlisting}

\endgroup

\end{minipage}
\end{figure}

\begin{table}[t]
\centering
\caption{\textbf{Query and label sets used for phishing content classification.}}
\label{tab:query_llm_methodology}
\footnotesize
\setlength{\tabcolsep}{5pt}
\renewcommand{\arraystretch}{1.12}

\resizebox{\columnwidth}{!}{%
\begin{tabular}{@{}p{1.5cm}p{1.4cm}p{5.2cm}@{}}
\toprule
\textbf{Input}
& \textbf{Dimension}
& \textbf{Prompt and Output Labels} \\
\midrule

\multirow{24}{*}{\shortstack{\textbf{Message text,} \\ \textbf{PDF, Image}}}
& \multirow{13}{*}{\textbf{Theme}}
& ``What type of theme does the email contain? Select the best choice from the following:
Account Security / Verification / Credential Alert;
Invoice / Payment;
Document Sharing;
Shipping / Delivery;
Internal Business Request / Boss Fraud;
Government Service;
Tech Support / Antivirus;
Reward;
Promotion / Product Offer;
Employment / Job Opportunity;
Extortion / Blackmail / Sextortion;
Romance / Personal Relationship;
Survey / Feedback;
Health / Medical;
Business Communication / Partnership Inquiry;
or Insufficient Content.'' \\

& \multirow{7}{*}{\textbf{CTA}}
& ``What behavioral mechanism or technical action is explicitly requested? Select the best choice from the following:
Navigate to a URL;
Perform Offline Communication;
Interact with Payload;
Direct Email Reply;
Forwarding;
Cryptocurrency Payment;
Multiple actions;
or No explicit action.'' \\

& \multirow{6}{*}{\textbf{Imp.}}
& ``Which specific organization, if any, does the content claim to represent? Select the best choice from the following:
Company / Private Sector;
Government / Public Sector;
or No Impersonation.
If impersonation is present, specify the organization name.'' \\

\midrule

\multirow{10}{*}{\textbf{Image}}
& \multirow{10}{*}{\textbf{Image Role}}
& ``What kind of functional role does this image have? Select the best choice from the following:
Document / Letter;
Logo / Brand Mark Only;
Button;
Advertisement / Promotional Banner;
QR Code Image;
Profile Photo / Human Photo;
Illustration / Hero Graphic;
Icon / Small UI Element;
ID Card;
Business Card;
Landscape / Scenery Photo;
Software Interface;
Device / Appliance / Instrument;
or No Discernible Role.'' \\

\bottomrule
\end{tabular}%
}

\end{table}

\section{Analysis of Cryptocurrency addresses}
\label{apx:analysis_crypto}
Of the 10,352 emails requesting a cryptocurrency payment, Bitcoin is the demanded currency in 85.5\% (8,852), followed by Litecoin (1,084; 10.5\%) and Monero (101; 1.0\%); a further 303 emails (2.9\%) offer the recipient a choice among multiple currencies, and Ethereum alone or an unspecified cryptocurrency account for the remaining 12 emails.
We exclude Monero from the address set because the protocol conceals recipient-side balances, making on-chain measurement of Monero addresses infeasible. Consequently, the 101 Monero-demanding emails contribute no addresses to our analysis.
73.5\% of these emails embed a syntactically valid wallet address directly in the message body, identified via per-currency address patterns with checksum verification.\footnote{Candidate addresses are
extracted with per-currency regular expressions and retained only if they pass checksum verification: Base58Check for Bitcoin/Litecoin legacy and Tron addresses, and BIP-173/350 polymod for bech32 addresses; Ethereum addresses are validated by format only.}
By querying the blockchain for transactions received by these addresses, we estimate the potential financial damage at approximately \$415K.
This suggests that our dataset captures attacks with measurable real-world impact.

The 7,606 addressed emails resolve to 2,539 unique wallet addresses, and reuse is heavily skewed: 438 addresses (17.2\%) appear in two or more emails and account for 75.3\% of all address occurrences, with the most frequent single address appearing in 465 emails (\autoref{tab:app_top_wallets} lists the five most reused address groups).
Note that the corpus is deduplicated (\autoref{sec:method:data:collection}); the 465 emails sharing this address are distinct messages rather than repeated submissions of a single email.

Moreover, address strategies differ across currencies: Bitcoin has 1,612 unique addresses across 6,531 address occurrences, whereas the Litecoin emails carry a nearly unique bech32 address per email (917 bech32 addresses over 1,070 occurrences). 
\autoref{tab:app_onchain} counts all 919 Litecoin addresses, including two legacy ones that contribute a further 294 occurrences.
To support reproducibility and follow-up analyses such as on-chain measurement, we release the full list of the 2,539 unique wallet addresses. The cryptocurrency-related measurement data used in this analysis are available in our anonymized artifact repository.\footnote{\url{https://anonymous.4open.science/r/anonymous_anynomous_anonoymous123/}}

\begin{table}[t]
  \centering
  \caption{\textbf{Top-5 Most Reused Wallet Address Groups.}
  Addresses in the same group co-occur in an identical set of emails;
  the groups are pairwise disjoint.
  Monero is excluded, as its balances are unobservable on-chain.}
  \label{tab:app_top_wallets}
  \footnotesize
  \setlength{\tabcolsep}{4pt}
  \resizebox{\columnwidth}{!}{%
  \begin{tabular}{cclr}
    \toprule
    \textbf{Group} & \textbf{Currency} & \textbf{Wallet address} & \textbf{Emails} \\
    \midrule
    G1 & BTC & \texttt{1Dhrsm6e4qzbMXKt4x84z6fd1btJSUcMJh}         & 465 \\
    \midrule
    \multirow{4}{*}{G2}
       & BTC & \texttt{1MK5DtGcAnfym3zmCCp5VqkG9iPpz6N2qZ}         & 289 \\
       & LTC & \texttt{LWHpbVsgqPRHJZYaf4dKKa5ZzSq4nQ9F1z}         & 289 \\
       & ETH & \texttt{0xb42b28fC4890aE540CFCb8e1992f3D27d5C5F8AF} & 289 \\
       & TRX & \texttt{TV8vcdP7UWPMhjGZZG5jQgtGC6N1cTCF1S}         & 289 \\
    \midrule
    G3 & BTC & \texttt{1G1zmqks1vd9V3SdxCY71Hv9C7rHBLQbCY}         & 264 \\
    G4 & BTC & \texttt{19S2iQNucjVZYSNLR8wWmJgx9Ucy4Uuv4K}         & 217 \\
    G5 & BTC & \texttt{1JM9twQwo5kEpicXhPvHiW5fpXJz6QXZQ2}         & 178 \\
    \bottomrule
  \end{tabular}%
  }
\end{table}

\PP{On-chain outcome}
We queried all 2,539 addresses for their complete transaction history on 2026-08-18, resolving Bitcoin and Litecoin through Esplora instances (\texttt{mempool.space}~\cite{mempoolapi:online}, \texttt{litecoinspace.org}~\cite{litecoinspaceapi:online}), Ethereum through Etherscan~\cite{etherscanapi:online}, and Tron through Tronscan~\cite{tronscanapi:online}; for Ethereum and Tron we additionally record ERC-20/TRC-20 inflows, as these chains are used to collect stablecoins rather than the native asset. 
Every address was resolved, and we cross-validated the figures against independent explorers (\texttt{blockchain.info}~\cite{blockchaininfoapi:online}, BlockCypher~\cite{blockcypherapi:online}), which reproduced them exactly. 
Each receipt is valued at the closing price of its own transaction date rather than at a single snapshot rate~\cite{coinbaseapi:online,binanceapi:online}.

\autoref{tab:app_onchain} reports the outcome. Only 215 of the 2,539 addresses (8.5\%) ever received a payment, across 816 incoming transactions worth \$414,979 in total, of which Bitcoin accounts for 87.7\%. 
Payments are highly concentrated: the median funded address collected \$1,396 while the largest single address collected \$19,463 over 208 transactions, and the ten largest addresses together hold 25.1\% of all proceeds. 
Receipts span 2018-10-27 to 2026-08-17---the latter falling after our collection window ended---indicating that at least part of this infrastructure remains active.

Two observations qualify the reuse pattern reported above. 
First, reuse does not predict revenue: across funded Bitcoin addresses, the rank correlation between how often an address appears and how much it received is negligible ($\rho = 0.02$, $n = 170$). G1, the most reused address, collected only \$1{,}801, whereas the five largest Litecoin receipts each came from an address appearing in a single email. 
Wide distribution instead raises the probability of being paid at all---funded Bitcoin addresses average 24.0 occurrences against 1.71 for unfunded ones. 
Second, offering victims a choice of currency did not pay off: in G2, which advertises one address on each of four chains, only the Bitcoin address was ever funded (0.0039\,BTC), while its Litecoin, Ethereum, and Tron counterparts remain empty.

\begin{table}[t]
  \centering
  \caption{\textbf{On-chain outcome per currency.}
  \emph{Occ.} counts address occurrences across emails (an email listing $k$
  addresses contributes $k$); \emph{Funded} counts addresses that received a
  non-zero amount. USD is valued at each transaction's own date.}
  \label{tab:app_onchain}
  \footnotesize
  \setlength{\tabcolsep}{4pt}
  \resizebox{\columnwidth}{!}{%
  \begin{tabular}{lrrrrrlr}
    \toprule
    \textbf{Cur.} & \textbf{Addr.} & \textbf{Occ.} & \textbf{Funded} & \textbf{\%} &
    \textbf{Txs} & \textbf{Received} & \textbf{USD} \\
    \midrule
    BTC          & 1,612 & 6,531 & 170 & 10.5 & 755 & 4.7674\,BTC            & 363,846 \\
    LTC          &   919 & 1,364 &  43 &  4.7 &  50 & 529.1615\,LTC          &  50,157 \\
    ETH          &     3 &   297 &   1 & 33.3 &   2 & 0.2716\,ETH            &     848 \\
    TRX          &     3 &   295 &   1 & 33.3 &   9 & 28.00\,TRX\,$+$\,118.20\,USDT & 128 \\
    P2SH$^\dagger$ &   2 &     2 &   0 &  0.0 &   0 & ---                    &     --- \\
    \midrule
    \textbf{Total} & \textbf{2,539} & \textbf{8,489} & \textbf{215} & \textbf{8.5} &
    \textbf{816} & --- & \textbf{414,979} \\
    \bottomrule
  \end{tabular}%
  }
  \\[2pt]
  \raggedright\footnotesize
  $^\dagger$Two \texttt{3\ldots} addresses whose version byte (\texttt{0x05}) is
  shared by Bitcoin and Litecoin; both are unused on either chain, so their
  currency cannot be determined.
\end{table}

\section{Phishing Image Examples}
\label{appxsec:phishing_image_role}

\begin{figure}[t]
    \centering
    \includegraphics[width=0.95\linewidth]{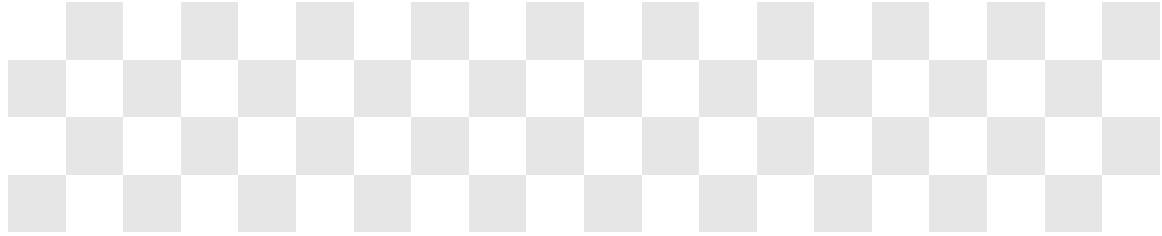}
    \caption{Example of an image categorized as ``No Discernible Role.''}
    \label{fig:appx_randomimage_example}
\end{figure}

\begin{figure}[t]
    \centering
    \includegraphics[width=0.20\linewidth]{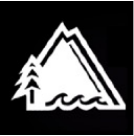}
    \caption{Example of an image categorized as ``Logo only'' (a generic emblem).}
    \label{fig:appx_logoimage_example_mountain}
\end{figure}

\begin{figure}[t]
    \centering
    \includegraphics[width=0.45\linewidth]{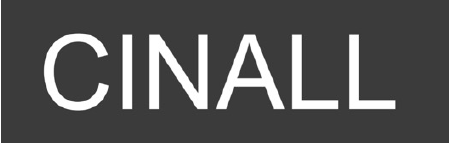}
    \caption{Example of an image categorized as ``Logo only'' (wordmark of an
unrecognized brand).}
    \label{fig:appx_logoimage_example_wordmark}
\end{figure}

\begin{figure}[t]
    \centering
    \includegraphics[width=\linewidth]{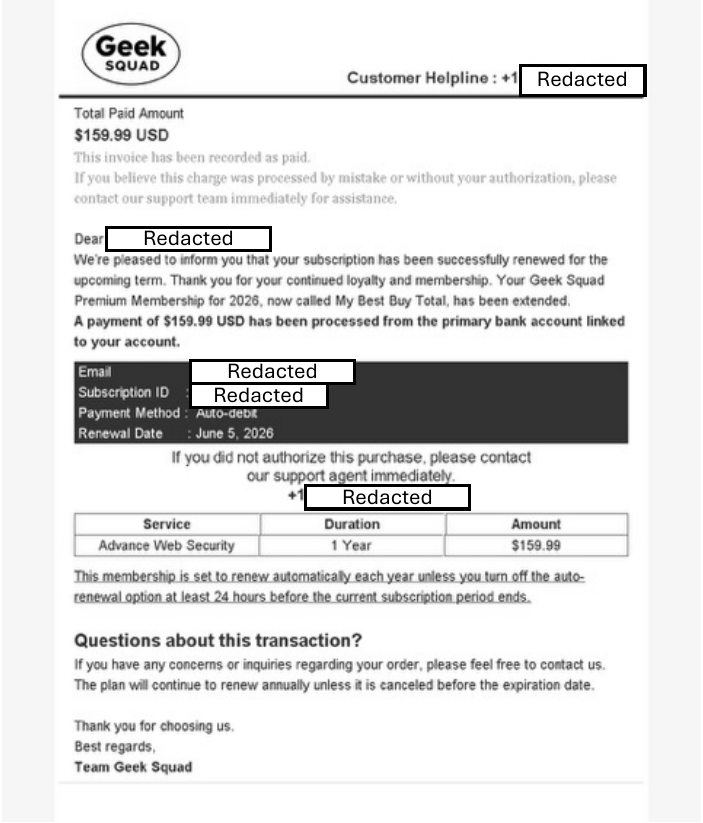}
    \caption{Example of an image categorized as ``Document.''}
    \label{fig:appx_logoimage_example_invoice}
\end{figure}

\begin{figure}[t]
    \centering
    \includegraphics[width=0.8\linewidth]{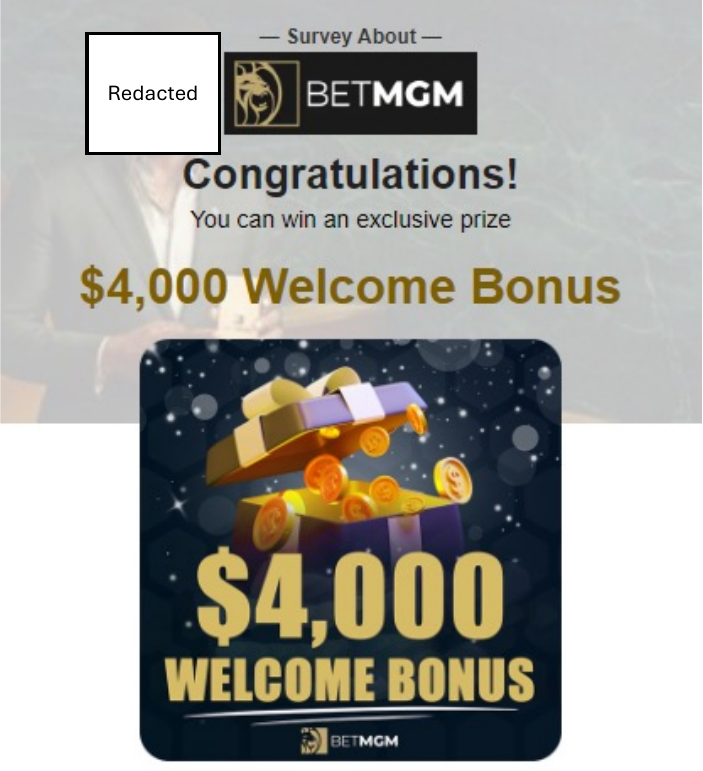}
    \caption{Example of an image categorized as ``Advertisement.''
 }
    \label{fig:appx_logoimage_example_advertisment}
\end{figure}

In this section, we present representative examples of each image role. \autoref{fig:appx_randomimage_example} shows an image categorized as ``No Discernible Role.'' \autoref{fig:appx_logoimage_example_mountain} and \autoref{fig:appx_logoimage_example_wordmark} show two forms of ``Logo Only:'' an emblem-style logo and a wordmark, respectively. \autoref{fig:appx_logoimage_example_invoice} shows a ``Document/Letter'' image, while \autoref{fig:appx_logoimage_example_advertisment} shows an ``Advertisement/Promotional Banner.''

\begin{table*}[t]
\caption{\textbf{Categories and Definitions Used for LLM-Based Classification.}}
\label{tab:llm_category_definitions}
\centering
\footnotesize
\begin{tabularx}{\textwidth}{
    @{}
    >{\raggedright\arraybackslash}p{1.6cm}
    >{\raggedright\arraybackslash}p{5.0cm}
    X
    @{}
}
\toprule
\textbf{Item} & \textbf{Category} & \textbf{Definition} \\
\midrule

\multirow{31}{*}{\textbf{Theme}}
& Account Security / Verification / Credential Alert
& Content that frames the lure around account access, identity, credentials, verification, or account control. \\

& Invoice / Payment
& Content primarily centered on billing, invoices, payments, transactions, refunds, charges, purchase orders, or receipts. \\

& Document Sharing
& Content centered on accessing, reviewing, opening, downloading, or signing a shared document, file, contract, report, or form, when document access itself is the primary lure. \\

& Shipping / Delivery
& Content centered on package delivery, shipment tracking, missed delivery, customs, or other delivery-related issues. \\

& Internal Business Request / Boss Fraud
& Content that frames the lure as a work-related request from a supposed internal authority, executive, manager, or colleague. \\

& Government Service
& Content primarily framed as an official government, regulatory, tax, court, or other public-service matter. \\

& Tech Support / Antivirus
& Content that frames the lure around technical support, device security, antivirus assistance, or system repair. \\

& Reward
& Content that frames the lure around receiving an unexpected or purportedly earned reward, prize, lottery winning, grant, compensation, or similar benefit. \\

& Promotion / Product Offer
& Content that frames the lure around a commercial product, service, sale, discount, subscription, marketing offer, or advertisement. \\

& Employment / Job Opportunity
& Content that frames the lure around recruitment, hiring, job applications, interviews, employment, payroll onboarding, or work opportunities. \\

& Extortion / Blackmail / Sextortion
& Content that frames the lure around threats, coercion, exposure, ransom, harm, or blackmail. \\

& Romance / Personal Relationship
& Content that frames the lure around romance, intimacy, friendship, dating, personal relationships, or emotional appeals. \\

& Survey / Feedback
& Content that frames the lure around collecting opinions, reviews, feedback, questionnaires, or survey responses. \\

& Health / Medical
& Content that frames the lure around health, medical care, healthcare services, insurance, medical benefits, pharmacy, or treatment. \\

& Business Communication / Partnership Inquiry
& Content that frames the lure as an external business inquiry, partnership proposal, procurement request, or vendor communication. \\

& Insufficient Content
& Content that lacks sufficient meaningful information to determine a phishing theme. \\

\midrule

\multirow{10}{*}{\textbf{CTA}}
& Navigate to a URL
& Requests that the recipient click a link or visit a URL. \\

& Perform Offline Communication
& Requests that the recipient call a phone number or send an SMS/text message. \\

& Interact with Payload
& Requests that the recipient download or open a file, install software, or review an attached or embedded document. \\

& Direct Email Reply
& Requests that the recipient compose a reply or send requested information through return email. \\

& Forwarding
& Requests that the recipient forward the email, message, or document to others. \\

& Cryptocurrency Payment
& Requests that the recipient send cryptocurrency to a wallet address or purchase cryptocurrency. \\

& Multiple actions
& Requests two or more distinct action types. \\

& No explicit action
& Contains no explicit action requested of the recipient. \\

\midrule

\multirow{3}{*}{\textbf{Impersonation}}
& Company / Private Sector
& Content that claims to represent a specific company or other private-sector organization. \\

& Government / Public Sector
& Content that claims to represent a specific government agency or other public-sector institution. \\

& No Impersonation
& Content that does not claim to represent a specific organization. \\

\midrule

\multirow{24}{*}{\textbf{Image Role}}
& Document / Letter
& A full or mostly complete document, invoice, receipt, contract, legal notice, shipping notice, tax notice, or business letter shown as an image, excluding ID cards and business cards. \\

& Logo / Brand Mark Only
& A logo or brand mark with no meaningful lure text or CTA. \\

& Button
& A small call-to-action element, such as a button or clickable-looking badge, whose primary role is to prompt user interaction. \\

& Advertisement / Promotional Banner
& A promotional image or banner centered on a product, service, sale, gambling offer, discount, or other marketing content. \\

& QR Code Image
& An image in which a QR code or barcode is the central visible element. \\

& Profile Photo / Human Photo
& A photograph of a person, such as a headshot, avatar, profile image, or other human-centered photo. \\

& Illustration / Hero Graphic
& A non-photographic illustration, decorative hero graphic, or conceptual drawing used as a prominent visual element. \\

& Icon / Small UI Element
& A small icon, badge, footer or social icon, app icon, or other minor UI or decorative element. \\

& ID Card
& A government-issued or official identification card, such as a driver's license, passport, national ID, or residence card, or a clear photo or scan of one. \\

& Business Card
& A personal or corporate business card showing a name, title, company, and contact details in a business-card layout. \\

& Landscape / Scenery Photo
& A photograph whose main subject is natural or outdoor scenery, such as landscapes, streets, or buildings, rather than a document, person, or product. \\

& Software Interface
& A screenshot or rendering of software, such as an application window, dashboard, console, terminal, code editor, or other program interface. \\

& Device / Appliance / Instrument
& A photograph or product image whose main subject is a physical device, appliance, electronic gadget, musical instrument, tool, or similar tangible product. \\

& No Discernible Role
& An image for which no meaningful functional role can be determined. \\
\bottomrule

\end{tabularx}

\end{table*}

\begin{table}[t]
\centering
\caption{\textbf{Themes, Call-to-Actions, and Impersonations in Document/Letter Images.}
$N$=79,559 images.}
\label{tab:image_document_composition}
\footnotesize
\setlength{\tabcolsep}{4pt}
\renewcommand{\arraystretch}{1.05}
\resizebox{\columnwidth}{!}{%
\begin{tabular}{c l r @{\hspace{10pt}} l r}
\toprule
& \textbf{Category} & \multicolumn{1}{c}{\textbf{Count (\%)}}
& \textbf{Category} & \multicolumn{1}{c}{\textbf{Count (\%)}} \\
\midrule

\multirow{8}{*}{\rotatebox[origin=c]{90}{\textbf{Theme}}}
& Invoice       & 57,801 (72.7\%) & Promotion             &   598 \phantom{0}(0.8\%) \\
& Tech support  &  4,798 \phantom{0}(6.0\%) & Health                &   512 \phantom{0}(0.6\%) \\
& Account       &  4,220 \phantom{0}(5.3\%) & Extortion             &   511 \phantom{0}(0.6\%) \\
& Government    &  3,334 \phantom{0}(4.2\%) & Document              &   450 \phantom{0}(0.6\%) \\
& Business      &  3,046 \phantom{0}(3.8\%) & Romance               &   213 \phantom{0}(0.3\%) \\
& Reward        &    745 \phantom{0}(0.9\%) & Internal Bus.         &   169 \phantom{0}(0.2\%) \\
& Shipping      &    691 \phantom{0}(0.9\%) & Survey                &   115 \phantom{0}(0.1\%) \\
& Employment    &    686 \phantom{0}(0.9\%) & Insufficient content  & 1,670 \phantom{0}(2.1\%) \\

\midrule

\multirow{4}{*}{\rotatebox[origin=c]{90}{\textbf{CTA}}}
& Offline comm.          & 60,680 (76.3\%) 
& Multiple               &    600 \phantom{0}(0.8\%) \\

& Navigate to URL        &  5,587 \phantom{0}(7.0\%) 
& Forwarding             &      4 \phantom{0}(0.0\%) \\

& Email reply            &  1,700 \phantom{0}(2.1\%) 
& No explicit action     &  9,810 (12.3\%) \\

& Payload                &  1,178 \phantom{0}(1.5\%) 
&                        & \\

\midrule

\multirow{2}{*}{\rotatebox[origin=c]{90}{\textbf{Imp.}}}
& Private Company        & 68,783 (86.5\%) 
& No impersonation       &  7,357 \phantom{0}(9.2\%) \\

& Government             &  3,419 \phantom{0}(4.3\%) 
&                        & \\

\midrule
\multicolumn{2}{l}{\textbf{Total images}} & & & \textbf{79,559} \\
\bottomrule
\end{tabular}%
}
\end{table}

\section{Analysis of Theme, CTA, and Impersonation by Image Role}
\label{appx:sec:image_role_content}

We report the theme, CTA, and impersonation distributions of the Document/Letter and Ad/Promo Banner images discussed in~\autoref{subsec:image_rq2} (see~\autoref{tab:image_document_composition} and~\autoref{tab:image_adpromo_composition}).
The two image roles exhibit sharply contrasting attack configurations.
Document/Letter images concentrate on an Invoice--Offline Communication
configuration: Invoice accounts for 72.7\% of themes and Offline
Communication for 76.3\% of CTAs, similar to the callback-oriented attack
model observed for PDFs (\autoref{subsec:pdfs_rq2}). In contrast, Ad/Promo
Banner images center on Promotion (78.1\%) with URL navigation (69.0\%)
as the dominant CTA, steering recipients to external websites. Despite
these divergent workflows, both roles rely predominantly on
private-company impersonation (86.5\% and 72.9\%, respectively).
Notably, ``No explicit action'' is more common in banners (23.7\%) than
in documents (12.3\%).

\begin{table}[t]
\centering
\caption{\textbf{Themes, Call-to-Actions, and Impersonations in Ad/Promo Banner Images.} $N$=23,677 images.}
\label{tab:image_adpromo_composition}
\footnotesize
\setlength{\tabcolsep}{4pt}
\renewcommand{\arraystretch}{1.05}
\resizebox{\columnwidth}{!}{%
\begin{tabular}{c l r @{\hspace{10pt}} l r}
\toprule
& \textbf{Category} & \multicolumn{1}{c}{\textbf{Count (\%)}}
& \textbf{Category} & \multicolumn{1}{c}{\textbf{Count (\%)}} \\
\midrule

\multirow{7}{*}{\rotatebox[origin=c]{90}{\textbf{Theme}}}
& Promotion     & 18,491 (78.1\%) & Account               &   164 \phantom{0}(0.7\%) \\
& Reward        &  2,444 (10.3\%) & Shipping              &    90 \phantom{0}(0.4\%) \\
& Survey        &    774 \phantom{0}(3.3\%) & Romance               &    73 \phantom{0}(0.3\%) \\
& Health        &    581 \phantom{0}(2.5\%) & Invoice               &    62 \phantom{0}(0.3\%) \\
& Employment    &    298 \phantom{0}(1.3\%) & Government            &    29 \phantom{0}(0.1\%) \\
& Business      &    269 \phantom{0}(1.1\%) & Document              &    19 \phantom{0}(0.1\%) \\
& Tech support  &    239 \phantom{0}(1.0\%) & Insufficient content  &   144 \phantom{0}(0.6\%) \\

\midrule

\multirow{3}{*}{\rotatebox[origin=c]{90}{\textbf{CTA}}}
& Navigate to URL & 16,344 (69.0\%) & Multiple            &   112 \phantom{0}(0.5\%) \\
& Offline comm.   &  1,560 \phantom{0}(6.6\%) & Email reply          &    46 \phantom{0}(0.2\%) \\
& Payload         &     10 \phantom{0}(0.0\%) & No explicit action  & 5,605 (23.7\%) \\

\midrule

\multirow{2}{*}{\rotatebox[origin=c]{90}{\textbf{Imp.}}}
& Private Company & 17,271 (72.9\%) &  No impersonation  & 6,077 (25.7\%) \\
& Government       &   329 \phantom{0}(1.4\%)              &            &  \\

\midrule
\multicolumn{2}{l}{\textbf{Total images}} & & & \textbf{23,677} \\
\bottomrule
\end{tabular}%
}
\end{table}

\section{Analysis of Theme, CTA, and Impersonation in Nazario}
\label{appx:sec:nazario_theme_cta_imp}

\begin{figure}[t]
    \centering

    \subfloat[Theme.]{
        \includegraphics[width=0.95\columnwidth]{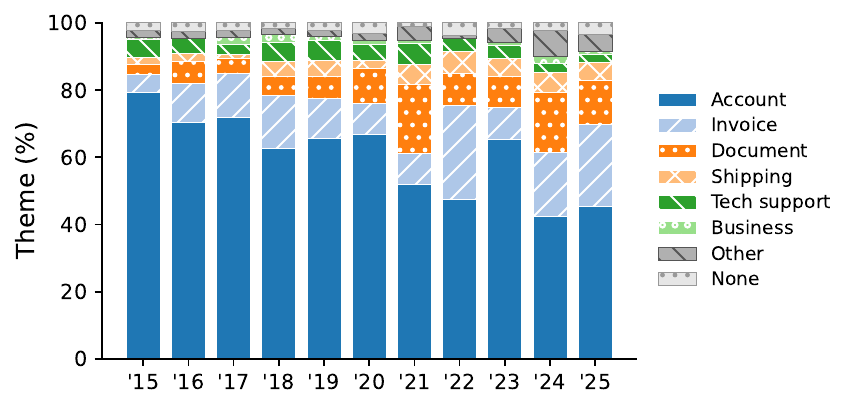}
        \label{fig:apwg_nazario_theme}
    }

    \subfloat[Call to Action.]{
        \includegraphics[width=0.95\columnwidth]{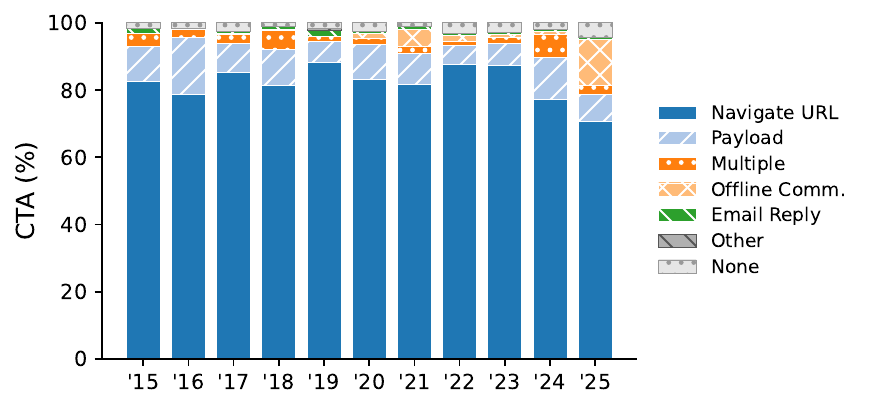}
        \label{fig:apwg_nazario_cta}
    }

    \subfloat[Impersonation.]{
        \includegraphics[width=0.95\columnwidth]{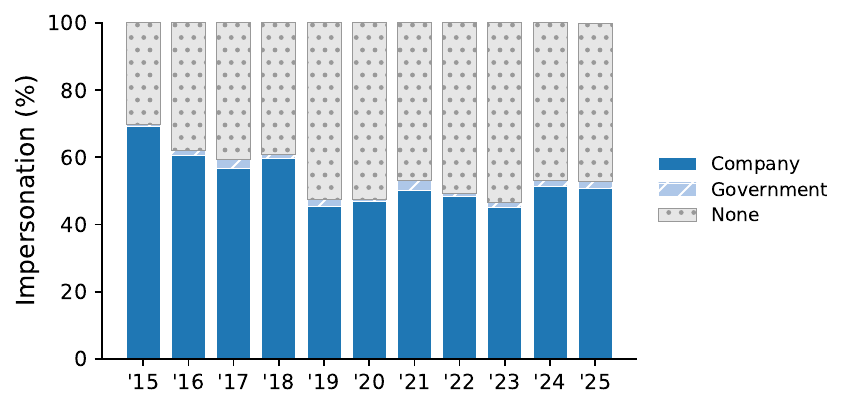}
        \label{fig:apwg_nazario_imp}
    }

    \caption{\textbf{Changes in phishing theme, call to action, and impersonation over time.}
    Yearly distributions in Nazario (2015--2025).
    Themes and impersonation patterns shift over time, while URL navigation remains the dominant call to action.}
    \label{fig:apwg_nazario_theme_cta_imp}
\end{figure}

We introduce the distribution of themes, CTAs, and impersonation in the Nazario dataset in~\autoref{fig:apwg_nazario_theme_cta_imp}.
Across 2015--2025, Account and Invoice are the only themes consistently
represented with substantial volume: Account remains the largest theme
in every year (42--79\%), while Invoice generally accounts for
10--25\%, motivating our selection of these two themes for longitudinal
comparison (\autoref{sec:researchquestion3_historical}). 

\section{Use of Generative AI}
We use GPT-5.1 as an analysis tool to classify phishing content at scale, as described in~\autoref{sec:method}. Specifically, the model is used to characterize message text, images, and PDFs along theme, call-to-action, and impersonation dimensions, as well as functional role for images only. We empirically evaluate the model against human-annotated samples before applying it to the full corpus. The authors are responsible for the design of the taxonomy, prompts, methodology, analysis, interpretation of results, and all claims in the paper.